%% file: main_v5_nat_comm.tex
\documentclass[superscriptaddress,aps,pra,twocolumn,showpacs,nofootinbib, longbibliography, notitlepage,floatfix]{revtex4-2}
\usepackage[utf8]{inputenc}
\usepackage[T1]{fontenc}
\usepackage{etex}
\usepackage{amsmath,amssymb,amsthm}
\usepackage[colorlinks=true,citecolor=blue,urlcolor=blue]{hyperref}
\usepackage[pdftex]{graphicx}
\usepackage{times,txfonts}
\usepackage{braket}
\usepackage{color}
\usepackage{natbib}
\usepackage{amsmath,blkarray}
\usepackage{mathtools}
\usepackage{ulem}
\usepackage{latexsym}
\usepackage{tabularx, booktabs}
\usepackage{graphics,epstopdf}
\usepackage{graphicx}
\usepackage{float}
\usepackage{amsfonts}
\usepackage{MnSymbol,graphicx}

\usepackage{natbib}

\usepackage{indentfirst}

\usepackage{tikz}
\usetikzlibrary{shapes,arrows,positioning}

\usepackage[ruled]{algorithm2e}
\usetikzlibrary{quantikz}
\usepackage{color,soul}

\newcommand{\be}{\begin{equation}}
\newcommand{\ee}{\end{equation}}
\newcommand{\ba}{\begin{eqnarray}}
\newcommand{\ea}{\end{eqnarray}}

\usepackage{multirow}
\usepackage{appendix}
\usepackage{url}
\usepackage{orcidlink}

\input{shortcut.tex}

\begin{document}
\title{{Noise-Assisted Feedback Control of Open Quantum Systems for Ground State Properties}}

\author{Kasturi Ranjan Swain\orcidlink{0009-0008-7227-3856}}
\email{{kasturi.swain@uni.lu}}
\affiliation{Department  of  Physics  and  Materials  Science,  University  of  Luxembourg,  L-1511  Luxembourg,  Luxembourg}
\author{Rajesh K. Malla\orcidlink{0000-0003-1052-3705}}
\email{{malla@aps.org}}
\affiliation{Physical Review Letters, American Physical Society, Hauppauge, New York 11788, USA}
\affiliation{Condensed Matter Physics and Materials Science Division, Brookhaven National Laboratory, Upton, New York 11973, USA}
\author{Adolfo del Campo\orcidlink{0000-0003-2219-2851}}                 
\affiliation{Department  of  Physics  and  Materials  Science,  University  of  Luxembourg,  L-1511  Luxembourg,  Luxembourg}
\affiliation{Donostia International Physics Center,  E-20018 San Sebasti\'an, Spain}

\begin{abstract}
Intrinsic noise in pre-fault-tolerant quantum devices poses a major challenge to the reliable realization of unitary dynamics in quantum algorithms and simulations. To address this, we present a method for simulating open quantum system dynamics on a quantum computer, including negative dissipation rates in the Gorini–Kossakowski–Sudarshan–Lindblad (GKSL) master equation. Our approach lies beyond the standard Markovian approximation, enabling the controlled study of non-Markovian processes within a quantum simulation framework. Using this method, we develop a quantum algorithm for calculating ground-state properties that exploits feedback-controlled, noise-assisted dynamics. In this scheme, Lyapunov-based feedback steers the system toward a target virtual state under engineered noise conditions. This framework offers a promising strategy for harnessing current quantum hardware and advancing robust control protocols based on open-system dynamics.
\end{abstract}
\maketitle 

Digital quantum simulation (DQS) provides a promising pathway for simulating quantum many-body systems using state-of-the-art quantum computers~\cite{Arute2019, Pan_PRL_2021}. However, inherent noise in quantum devices remains a significant obstacle to performing meaningful computations, especially when aiming to demonstrate a quantum advantage over classical computers. Pre-fault-tolerant quantum devices commonly experience dissipation, decay, and decoherence, significantly limiting the circuit depth and the manageable system size~\cite{Bharti_RevModPhys_2022}. In parallel with hardware advances that reduce environmental effects, extensive efforts are underway to address noise through quantum error mitigation~\cite{Cai_QEM_2023}, quantum error correction~\cite{PhysRevA_surface_code}, and noise suppression methods. Ongoing theoretical and experimental advances are jointly driving progress toward realizing quantum advantage.

DQS in quantum devices typically requires Trotterization of the system dynamics into native quantum gates. However, due to intrinsic noise, the dynamics of closed systems realized in actual quantum hardware often resemble that of open quantum systems (OQS). Several theoretical and experimental studies have proposed harnessing noise as a resource for various applications, including thermal and ground-state preparation~\cite{Terhal_pra_2000, kraus_pra_2008, Verstraete2009, Kapit_2017, science_dissipation}, quantum information processing~\cite{Harrington2022_nat_phys}, and simulation of open quantum systems~\cite{PhysRevA.108.062424, Pleino_PRX_2023}. Such efforts are crucial, as error correction generally incurs substantial overhead in physical qubits and gates, making its practical implementation challenging~\cite{PhysRevA_surface_code}.

To simulate open quantum systems, rather than conventional isolated many-body systems, a variety of strategies have been proposed ~\cite{Nori_pra_2011, Solano_Pra_2016, PRXQuantum.5.020332, Delgado_Granados2025}, including simulating the Lindblad master equation via imaginary-time evolution~\cite{PRXQuantum_Motta_2022}, the Kraus operator formalism~\cite{Hu2020, Mazziotti_PRL_2021}, Trotterization methods~\cite{PhysRevLett.127.020504}, and variational algorithms~\cite{PRR_VQE_NESS_2020_Fujii, 2020_prl_Endo}. However, these methods often encounter scalability issues because of the need for ancillary qubits or mid-circuit measurements. As a result, stochastic approaches present a systematic solution to overcome scalability limitations~\cite{Aurelia_adc_2017, Narang_PRR_2021, peng2024quantumtrajectoryinspiredlindbladiansimulation, Sandro_Donadi_PRR_2024, borras2024quantumalgorithmsimulatelindblad, liu2024simulationopenquantumsystems, huang2025robustvariationalquantumsimulation}. Furthermore, accurately modeling environmental noise frequently involves information flow from the environment back into the system, violating the Markovian approximation. The framework of non-Markovian open quantum systems thus offers a more realistic description and deeper insight into dissipation and decoherence processes~\cite{Zoller_book_noise}. Therefore, algorithms that incorporate non-Markovian dynamics are central to the realistic simulation of open quantum systems ~\cite{Modi_PRXQuantum_2021, cai_non_markov_2025}.

Here we propose a method to simulate a class of non-Markovian time-dependent dynamics of an OQS in a quantum circuit via the pseudo-Lindblad quantum trajectory (PLQT) approach~\cite{Eckardt_PRL}. This is motivated by the Monte Carlo wavefunction method and does not require any additional ancillary qubits or an expanded Hilbert space~\cite{Dalibard_etal_prl, Carmichael1993Open}. With this approach, an arbitrary noisy process in the Lindblad form can be implemented on an ideal quantum circuit using any quantum computing software development kit, such as the AerSimulator of Qiskit~\cite{Qiskit2021}. However, in the quantum trajectory approach, the system evolves through pure states, a condition that is generally unattainable in experiments. For experimental implementation, we demonstrate a method for using noise as a resource to simulate non-Markovian time-dependent dynamics with the help of quantum error mitigation techniques such as the quasiprobability decomposition (QPD) method~\cite{Gambetta2017PRL, Cai_QEM_2023}. A critical step in our algorithm is to characterize intrinsic noise in each circuit layer using randomized compiling and cycle error reconstruction methods~\cite{EmersonPRA2016, Flammia2020, EmersonPRX2021}. Error mitigation protocols allow one to compute the expectation values of quantum observables and may provide a path toward the quantum advantage~\cite{Blatt_nat_comm_2019, vandenBerg2023_natphys, Kim2023, aharonov2025importanceerrormitigationquantum, zimborás2025mythsquantumcomputationfault}. The motivation of this work aligns with that of improving the capabilities of near-term quantum hardware to perform useful tasks.

We integrate our scheme to simulate non-Markovian OQS dynamics with feedback-based quantum algorithms (FQAs), introducing the noise-assisted feedback-based quantum algorithm (NAFQA). FQAs were originally proposed to find solutions to combinatorial optimization problems~\cite{SarovarPRL2022, Magann_PRA_2022}, with broad applications in industry and academia, ranging from job-shop scheduling to drug discovery. However, experimental implementations often fail to find optimal solutions due to the intrinsic noise of the device~\cite{SarovarPRL2022, RKMalla_PRR_2024}. We apply NAFQA to the max-cut and the transverse field Ising models. Our results demonstrate that feedback-based OQS dynamics with intrinsic noise converges more rapidly to the virtual ground state than closed-system dynamics, as the presence of noise enforces a negative energy gradient at each step, thereby accelerating population transfer to low-energy states. Similar to other error mitigation strategies, our approach has an exponential overhead to some parameters~\cite{Takagi2022}. Its applicability is limited to physical scenarios with a low number of decay channels and does not constitute an alternative to quantum error correction.

Although we primarily focus on the DQS here, our approach is not limited to gate-based quantum computers and can also be realized in analog settings. The QPD assumes a simplified noise channel that is modeled before or after the Trotterized layer of a quantum circuit. In contrast, analog simulations do not need discretization, and our non-Markovian dynamics protocol can be realized with continuous pulses of recovery operations~\cite{Jinzhao_Sun_PhysRevApplied_2021}. Our approach may also find applications in error correction, particularly in autonomous quantum error correction, where dissipation helps stabilize trapped-ion qubits~\cite{Reiter_Zoller_nat_comm_2017}.\\

\noindent \textbf{\large Results}\\
Our protocol follows a two-step approach. The first step aims to characterize and control the noise associated with each layer by constructing a noise map that integrates noise characterization techniques with the PLQT or the QPD method. In the second step, we introduce feedback-based control parameters using quantum Lyapunov control to accelerate convergence toward the virtual many-body ground states. We emphasize that our approach enables the determination of ground-state properties—such as short- and long-range correlators, dipole moments, and related observables—at the end of the protocol, rather than the true ground state itself.\\

\noindent \textbf{Non-Markovian dynamics.}~
The implementation of quantum circuits in DQS inherently involves noise, which can arise from various sources, such as cross-talk, dephasing, energy relaxation, charge noise, etc. To learn the noise, the errors can be first converted to stochastic Pauli errors by randomly applying Paulis before and after noisy gates with a technique known as randomized compiling or Pauli twirling~\cite{EmersonPRA2016, EmersonPRX2021}. Then, the noise associated with each layer of the ideal unitary operator $U$ can be modeled by introducing a stochastic Pauli noise map $\Lambda$ along with the ideal map $\mathcal{U}$. The calligraphic symbol $\mathcal{U}$ denotes a quantum channel corresponding to a unitary $U$ and its action on the density matrix is defined as $\mathcal{U}: \rho \rightarrow U \rho U^{\dagger}$, where $U$ consists of single- and two-qubit Clifford gates. The stochastic noise map $\Lambda$ corresponds to a mixed unitary quantum channel and is expressed as 
\begin{equation}
    \rho \rightarrow \Lambda(\rho) = \sum_{P \in \{I,X,Y,Z\}^{\otimes N}} \lambda_P P \rho P^{\dagger},\label{gen_pauli_channel}
\end{equation}
where $I, X, Y, Z$ denotes the Pauli matrices, and $\lambda_P$ is the relative error probability associated with the Pauli string $P$ that satisfies the condition $\sum_P \lambda_P = 1$. These error probabilities $\{ \lambda \}$ of noisy intermediate-scale quantum (NISQ) devices can be learned efficiently, as demonstrated in recent experiments~\cite{Blatt_nat_comm_2019, vandenBerg2023_natphys, Kim2023, fischer2024dynamicalsimulationsmanybodyquantum}.

Once the noise is learned, we add an additional noise map $\Lambda_s'$ via the PLQT or the QPD method~\cite{Eckardt_PRL, Gambetta2017PRL}. Here, we use the subscript `$s$' for the $s^{\text{th}}$ Trotter step. The combination of these maps simulates effective non-Markovian, time-dependent dynamics, depicted as
\begin{equation}
% \centering
\begin{tikzpicture}
    % Define nodes
    \node (rho) {\large $\rho$};
    \node[draw, minimum size=6mm, right=0.5cm of rho] (E1) {\large $\Lambda_s'$};
    \node[draw, minimum size=6mm, right=0.5cm of E1] (G) {\large $\Lambda_s$};
    \node[draw, minimum size=6mm, right=0.5cm of G] (E2) {\large $\mathcal{U}_s$};
    \node[right=0.5cm of E2] (rho') {\large $\rho'$ ~.};

    % Draw lines
    \draw[-] (rho) -- (E1);
    \draw[-] (E1) -- (G);
    \draw[-] (G) -- (E2);
    \draw[-] (E2) -- (rho');

    \node[draw, dashed, fit=(G) (E2), inner sep=5pt] (box) {};
    \node[below=0pt of box] {Noise Model};\label{non_Markovian-ckt}
\end{tikzpicture}
\end{equation}
The resulting noisy operation $\mathcal {F}_s = \Lambda_s \circ \Lambda_s'$,  with `$\circ$' denoting the composition of the maps,  captures the targeted non-Markovian dynamics, which can be associated with the Gorini–Kossakowski–Sudarshan–Lindblad (GKSL) master equation~\cite{Breuer_book} with negative decay rates. Note that Eq.~\eqref{non_Markovian-ckt} can be expressed as $\rho(t+ dt) = \mathcal{\tilde{U}}_s \mathcal {F}_s (\rho(t))$, where the superopertaor $\mathcal {F}_s$ represents the noisy dynamics for a small time interval of $dt$. Our method can be understood as bringing the noise source of NISQ devices into contact with an additional bath, with the combined effect designed as a resource for quantum computation. 

We emphasize that although negative error rates represent unphysical noisy channels, they can be included in the quantum circuit through techniques such as the PLQT or the QPD approach, with the correct classical post-processing~\cite{Donvil2022, Eckardt_PRL, vandenBerg2023_natphys, Gambetta2017PRL}. For an arbitrary rate $\Gamma$, the trajectories corresponding to the stochastic unraveling of the GKSL master equation evolve independently. The PLQT approach expands the system's Hilbert space with a single classical bit $s(t) = \pm1$, $\{\ket{\psi(t)}\} \rightarrow \{\ket{\psi(t)}, s(t)\}$. Starting from an initial pure state, stochastic unraveling of the state vectors leads to the final state as an ensemble average over pure states, i.e., $\rho(t) = \overline{s(t) \ket{\psi(t)} \bra{\psi(t)}} $~\cite{Dalibard_etal_prl, Carmichael1993Open}. Moreover, any property of interest can be found with knowledge of the state as $\braket{O} = \text{Tr}(\rho O)$; in this work, we focus on the ground-state energy. The classical post-processing handles the sign problem by storing the sign $s(t)$ at each step for every trajectory. We refer to Methods for details of noise learning, the PLQT approach, and sampling complexity.\\

\begin{figure*}[t!]
\includegraphics[scale=0.28]{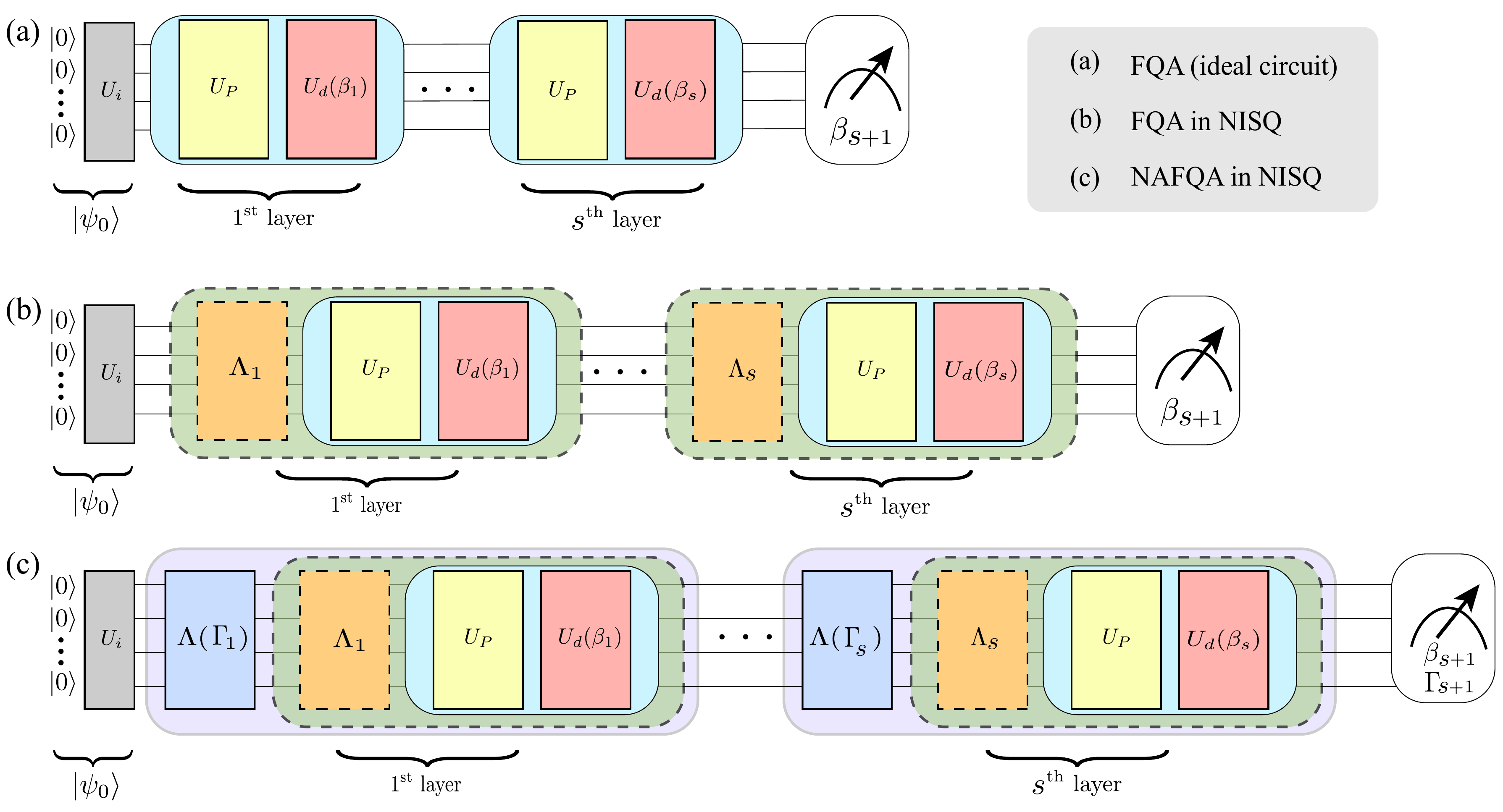} 
    \caption{\textbf{Schematic diagram of FQA in ideal, FQA in NISQ, and NAFQA in NISQ devices.}~(a) Ideal implementation of FQAs in a noiseless quantum computer consisting of $s$ Trotter layers. (b) Actual implementation of FQAs in NISQ devices where the ideal layer containing $U_p$ and $U_d(\beta_i)$ is interleaved with the inherent noise channel $\Lambda_s$. The dashed line represents the intrinsic device noise, over which the user generally has no control. (c) Our NAFQA algorithm is demonstrated to utilize noise to drive an OQS simulation by introducing an additional noise map $\Lambda(\Gamma_s)$ based on the feedback law. Note that the noise channels $\Lambda$ represent non-unitary evolutions and can be obtained via the stochastic unravelings of the GKSL master equation.} \label{Fig: Schematic_plot}
\end{figure*}

\noindent \textbf{Quantum Lyapunov controlled feedback-based quantum algorithm.}~Quantum Lyapunov control (QLC) is the founding principle behind recently developed FQAs designed for closed many-body systems. The QLC is a robust method that steers the dynamics of a quantum system towards a desired target state by tuning the Hamiltonian parameters. This is particularly applicable to finding approximate solutions to combinatorial problems or the ground state of many-body quantum Hamiltonians~\cite{SarovarPRL2022, Magann_PRA_2022, Dalcco_2024, RKMalla_PRR_2024, Larsen_prr, Shao_PhysRevA_2024}. For a closed quantum system described by a problem Hamiltonian $H_p$, the density matrix $\rho$ evolves according to the von Neumann equation $d \rho/dt = -i [H(t), \rho]$, where the time-dependent Hamiltonian $H(t) = H_p + \beta(t) H_d$, includes a control parameter $\beta(t)$ and a control Hamiltonian $H_d$. To ensure convergence towards the ground state, one should impose the Lyapunov condition $\frac{d}{dt} \braket{H_p} \leq 0,~ \forall t \geq 0$, which leads to
\begin{equation}
    \frac{d}{dt} \braket{H_p} = \beta(t) \braket{\psi(t) | i[H_d, H_p] | \psi(t)} = \beta(t) A(t) \leq 0,\label{closed_system}
\end{equation}
where $A(t) = \braket{\psi(t) | i[H_d, H_p] | \psi(t)}$. The above inequality can be satisfied by choosing an appropriate $\beta(t)$, and a convenient choice is $\beta(t) = - A(t)$. This ensures a negative energy gradient at all time steps. FQA employs a Trotterized version of QLC, where the evolution time $t$ corresponds to discrete layers in a quantum circuit. A lower bound for $\beta(t)$ is derived in the Supplementary Information (SI) and is given by $\beta(t) \geq - \frac{\sqrt{||H_p||} ~ \sqrt{||H_d||}}{2}$, where $||\cdot||$ is the semi-norm of an operator. Similarly, a bound for the Trotter step $dt$ can be obtained in terms of the spectral norms of $H_p$ and $H_d$ as, $dt \leq \frac{1}{4 ||H_p||_2 ||H_d||^2_2}$~\cite{Larsen_prr}.

\begin{figure*}[t!]
    \begin{tabular}{c c}
    \centering
    \includegraphics[scale=0.58]{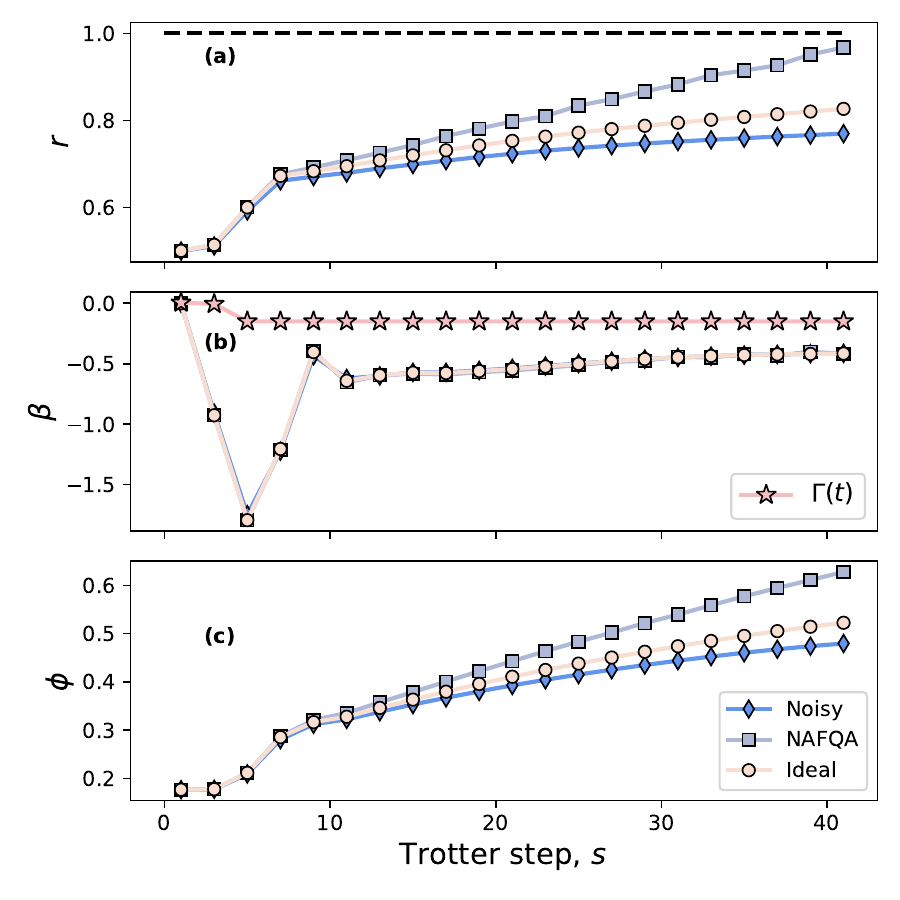} & 
    \includegraphics[scale=0.347]{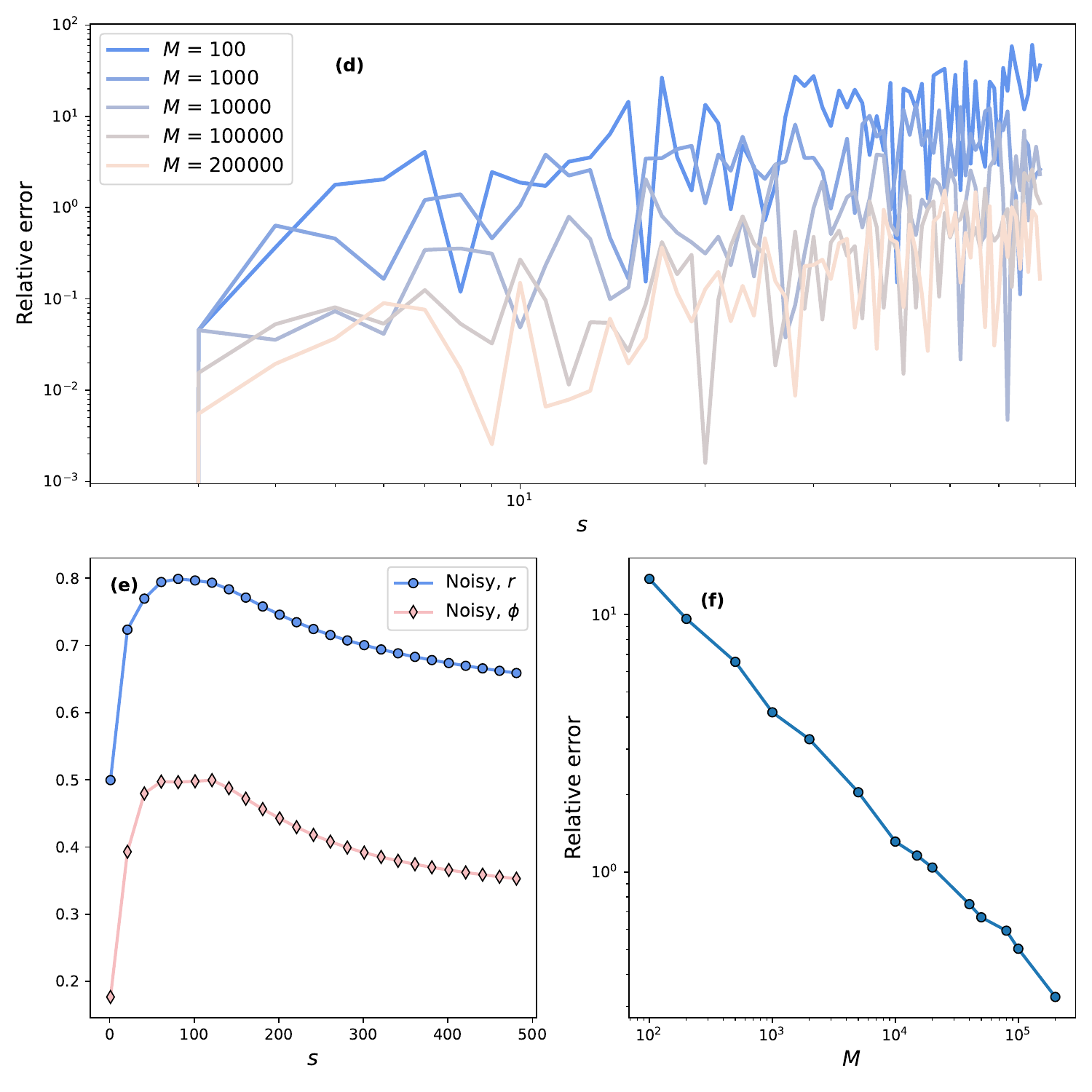}
    \end{tabular}
    \caption{\textbf{Numerical results for the 5-qubit Maxcut problem.}
    (a) The approximation ratio $r$, (b) the control terms $\beta(t)$ and $\Gamma(t)$, (c) the success probability $\phi$ as a function of the Trotter step with a fixed time step $dt = 0.07$. The inset in (b) shows $\Gamma(t)$ for the Pauli error term $IIYII$, whereas all the other Pauli terms' error probabilities are set to zero.  (d) Relative error as a function of the number of layers in the double logarithm scale with several numbers of trajectories, $M$. (e) $r$ and $\phi$ at large times demonstrate the non-monotonic behavior. (f) Time averaged relative error as a function of $M$. }\label{one_decay_op}
\end{figure*}
% ntraj = 100000
% s = 41
We note that the statistical errors in estimating the expectation values do not accumulate over time, despite some uncertainties in determining $\beta(t)$ in each run. The overall uncertainty in statistical errors scales as $\frac{1}{\sqrt{M}}$, where $M$ is the number of repetitions for each measurement~\cite{RKMalla_PRR_2024}.\\

\noindent  \textbf{Noise-assisted feedback-based approach.}~The intrinsic noise of NISQ devices can be modeled using the Markovian theory of OQS~\cite{vandenBerg2023_natphys, Ferracin2024efficiently}, and can be written in the canonical form of the GKSL master equation~\cite{Breuer_book},
\begin{equation}
    \frac{d \rho}{dt} = -i [H, \rho] + \mathcal{D}_{\text{int}} [\rho]\label{rho_evolution}.
\end{equation}
The dissipator $\mathcal{D}_{\text{int}} [\rho]$ accounts for the system-environment coupling and is expressed as $\mathcal{D}_{\text{int}} [\rho] = \sum_i (L_i \rho L_i^\dagger - \frac{1}{2} L_i^\dagger L_i \rho - \frac{1}{2} \rho L_i^\dagger L_i )$, where the dissipation rates are absorbed into the Lindblad operators $L_i$.

To integrate the QLC into the OQS framework, we require that the rate of change of energy corresponding to the problem Hamiltonian $H_p$ must be negative,  
\begin{equation}
    \frac{d}{dt} \braket{H_p} = \beta(t)  \braket{i[H_d, H_p]} + \text{Tr}(\mathcal{D}_{\text{int}} [\rho] H_p) \leq 0. \label{open_system}
\end{equation}
The first term is similar to Eq.~\eqref{closed_system}, whereas the second term can be expanded as
\begin{equation}
\begin{split}
    \text{Tr}(\mathcal{D}_{\text{int}} [\rho] H_p) & = \sum_i \braket{ L_i^\dagger H_p L_i } - \sum_i \frac{1}{2} \braket{ \{H_p, L_i^\dagger L_i\} }\\ & =\braket{\mathcal{D_{\text{int}}^{\dagger}}[H_p]},\label{superoperator_D}
\end{split}
\end{equation}
where $\{{A}, {B} \}$ denotes the anti-commutator of two operators ${A}$ and ${B}$.
Thus, it reduces to the expectation value with respect to $\rho$ of the adjoint of the dissipator acting on $H_p$.  In the case of a pure dephasing channel, with the Lindblad operator $L \propto H$, Eq.~\eqref{superoperator_D} vanishes~\cite{wiseman2010quantum}, and we arrive at a Lyapunov condition that is similar to that introduced for closed quantum systems in Eq.~\eqref{closed_system}.

Randomized compiling or Pauli twirling have been proposed and experimentally realized to simplify the device noise as stochastic Pauli noise channels~\cite{Blatt_nat_comm_2019, vandenBerg2023_natphys, Kim2023}, as discussed in Eq.~\eqref{gen_pauli_channel}. Using this technique, the dissipator takes the form~\cite{EmersonPRA2016, Blatt_nat_comm_2019} 
\begin{equation}
    \mathcal{D}_{\text{int}} [\rho] = \sum_{k\in \mathcal{K}} \lambda_k (P_k \rho P_k^{\dagger} - \rho),\label{pauli_channel}
\end{equation}
where $L_k = \sqrt{\lambda_k} P_k$, and $\mathcal{K}$ is the selected set of local Pauli operators $\{P_k\}_{k\in\mathcal{K}}$ with associated error probabilities $\{\lambda_k\}_{k\in\mathcal{K}}$. The correlated noise in large quantum circuits can be modeled as a sparse channel and is given by~\cite{vandenBerg2023_natphys}
\begin{equation}
    \Lambda(\rho) = \text{exp}[\mathcal{D}_{\text{int}}](\rho) = 
    \prod_{k\in\mathcal{K}} \left( \omega_k \cdot + (1-\omega_k) P_k \cdot P_k^{\dagger} \right) \rho, \label{noise_channel_Lambda1}
\end{equation}
where $\omega_k = (1+e^{-2\lambda_k})/2$.  Here, $\mathcal{K}$ represents the local noise interactions of the quantum device, which is governed by the qubit-coupling topology. 
We assume on physical grounds a sparse model, in which the number of Pauli noise operators scales polynomially in the number of qubits $N$, i.e., $|\mathcal{K}| \ll 4^N-1$. 
In addition, although accurately characterizing noise remains an active area of research~\cite{PRXQuantum_Zoller_2022, Chen2023}, we further assume that the noise in NISQ devices can be reliably learned. Using Eq.~\eqref{pauli_channel},  Eq.~\eqref{open_system} becomes
\begin{equation}
\begin{split}
    \frac{d}{dt} \braket{H_p} & = \beta(t)  \braket{i[H_d, H_p]} \\
    & + \sum_{k\in \mathcal{K}} \lambda_k \left(\braket{ P_k^\dagger H_p P_k } - \braket{H_p}\right) \leq 0. \label{lyapunov_lambda}
\end{split}
\end{equation}
The second term in Eq.~\eqref{lyapunov_lambda} can be positive and is responsible for the deviation from the monotonic decrease in $E_p$ and may explain the non-monotonic behavior of the energy as a function of the circuit depth in prior quantum hardware simulations~\cite{SarovarPRL2022, RKMalla_PRR_2024}. 

\begin{figure}[t!]
    \centering \includegraphics[scale=0.33]{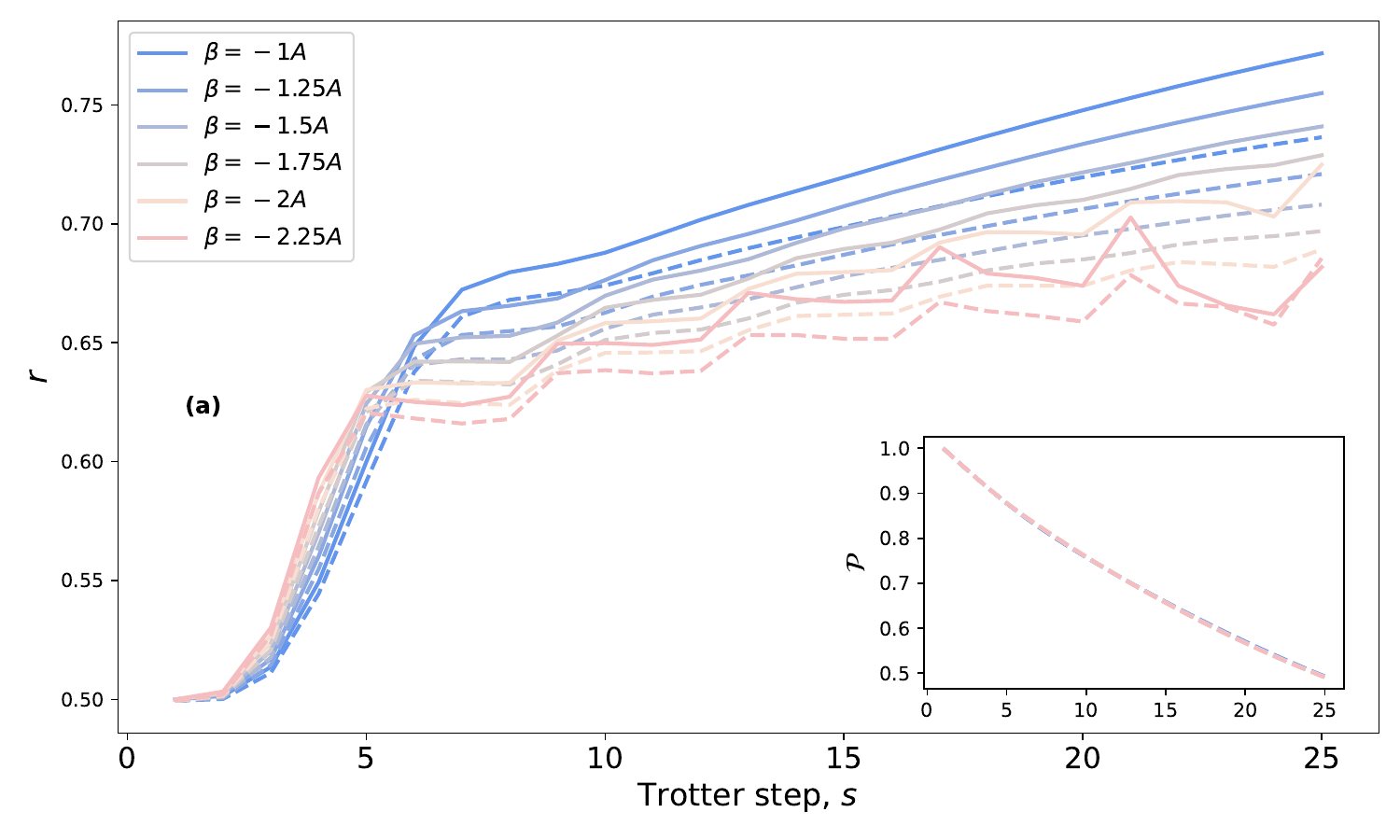} \\
    \centering \includegraphics[scale=0.33]{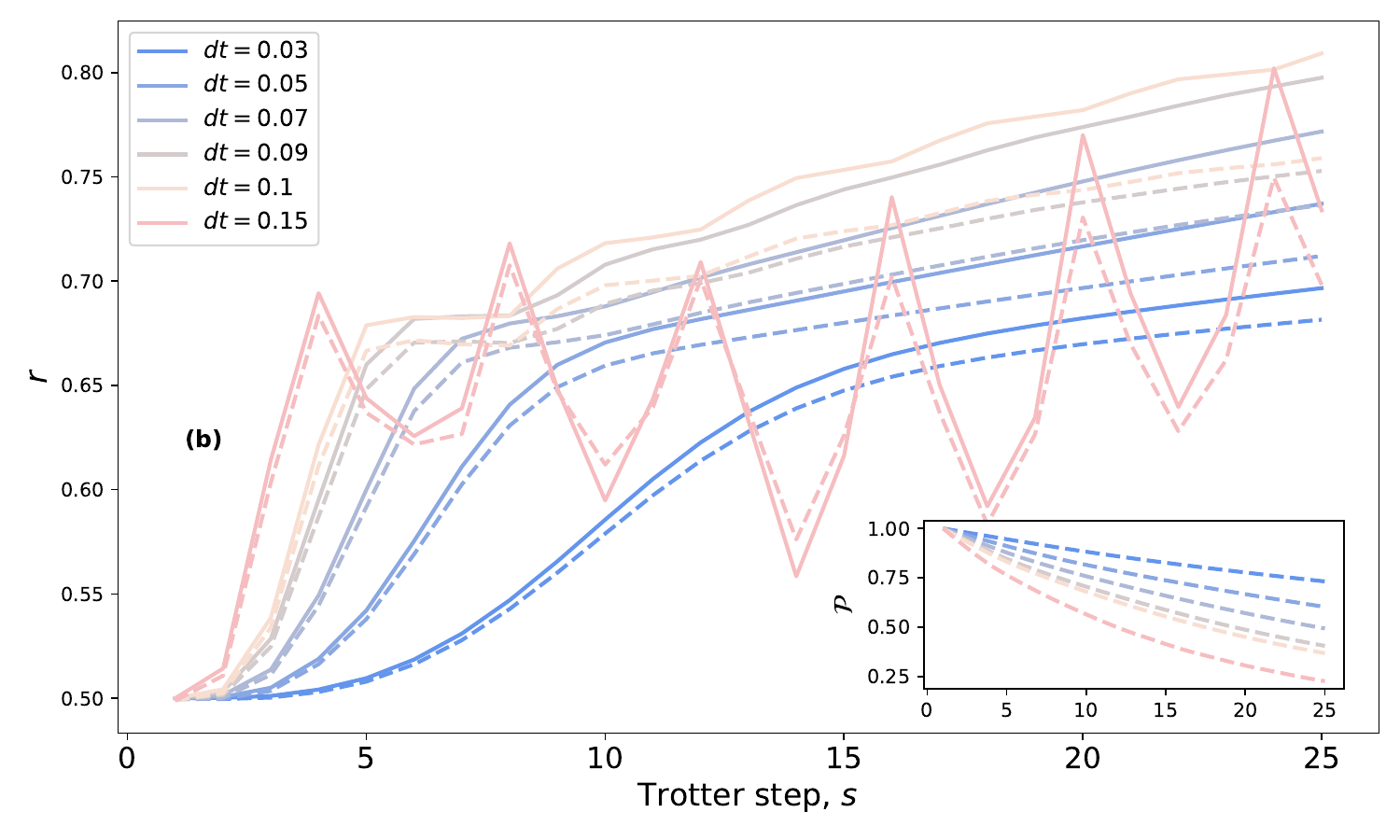} 
    \caption{\textbf{Performance of FQA in the closed and open quantum system framework.} Approximation ratio $r$ for the 5-qubit Maxcut problem with varying the (a) control field $\beta$ and (b) Trotter-step $dt$. Panel (a) is shown for the fixed Trotter step $dt = 0.07$, whereas panel (b) is illustrated for $\beta = - A$. The dotted and solid lines represent the noisy and ideal evolutions, respectively. The inset shows the purity $\mathcal{P}$ of the noisy evolutions as a function of Trotter steps.}\label{beta_change_purity}
\end{figure}

One natural question arises: can making 
$\beta(t)$ more negative accelerate convergence to the ground state?
The answer is no. The intrinsic noise present in quantum devices effectively heats the system, resulting in a mixed final state when FQAs are run on NISQ devices. Consequently, the true ground state cannot be reached in this manner. Similar questions can also be asked to increase the Trotter step $dt$ (instead of decreasing $\beta(t)$); however, the answers remain the same. In addition, the parameters cannot be arbitrarily large or small, and there is a bound for the $\beta(t)$ and $dt$ values that will depend on the choices of $H_p$, $H_d$~\cite{Magann_PRA_2022, Larsen_prr}; see SI. We demonstrate these statements numerically in Fig.~\ref{beta_change_purity} and discuss these aspects in more detail later in the paper. This observation motivates the development of an improved cooling strategy, which we outline below.

Alternatively, one may view this as suppressing the deviations (the second term in Eq.~\eqref{lyapunov_lambda}) by dynamically adjusting the error probabilities over time, or equivalently, across layers of the quantum circuit. Here, we introduce an additional Lyapunov control that depends on the modified error probabilities $\{\Gamma_k(t)\}_{k\in\mathcal{K}}$, such that $\Gamma_k(t) = - C_k(t)$, where $C_k(t) = \left(\braket{ P_k^\dagger H_p P_k } - \braket{H_p}\right)_t$. For consistency, we keep the standard Lyapunov condition on $\beta(t)$, i.e., $\beta(t) = -A(t)$. 

The time-local master equation corresponding to the layer-dependent rates $\{\Gamma\}$ defined above can be expressed as 
\begin{equation}
    \frac{d \rho(t)}{dt} = -i [H(t), \rho(t)] + \sum_{k\in \mathcal{K}} \Gamma_k(t) (P_k \rho(t) P_k^{\dagger} - \rho(t)),\label{na--qlc}
\end{equation}
where the sign of $\Gamma_k(t)$ determines whether the noise map is Markovian or non-Markovian. The bounds for the decay rates $\Gamma_k(t)$ are derived in the SI in terms of purity and fidelity between the initial and final states. However, one can obtain Eq.~\eqref{na--qlc} in a quantum device following Eq.~\eqref{non_Markovian-ckt}. The corresponding master equation reads
\begin{equation}
    \frac{d \rho(t)}{dt} = -i [H(t), \rho(t)] + \mathcal{D}_{\text{int}} [\rho(t)] + \mathcal{D}_{\text{eng}} [\rho(t)],\label{NAFQA-qlc}
\end{equation}
where the engineered dissipator takes the form $\mathcal{D}_{\text{eng}} [\rho(t)] = \sum_{k\in \mathcal{K}} \nu_k(t) (P_k \rho(t) P_k^{\dagger} - \rho)$, with time-dependent rates $\nu_k(t)$ that leads to $\Gamma_k(t) = \lambda_k + \nu_k(t)$. The overhead cost for this case, as derived in Ref.~\cite{Enod_PhysRevA_2021}, is given by
\begin{equation}
    \mathcal{N} (T) = \text{exp}\left[\sum_{k\in \mathcal{K}} \int_0^T\left( |\nu_k(t)| + \nu_k(t) \right) dt\right],\label{error_mitigation_cost}
\end{equation}
which can also be related to a measure of non-Markovianity. According to the Hoeffding inequality, the number of samples $M$ that ensure an average expectation value with error $\delta$ and probability $1-\epsilon$ is given by $M \propto \mathcal{N} ^2 \frac{\text{log}(1/\epsilon)}{{\delta^2}}$~\cite{Jinzhao_Sun_PhysRevApplied_2021}. As a result, the sampling overhead scales as $\mathcal{N} ^2$, requiring an increased cost related to the negativity of the rates to guarantee a fixed precision. We note that the overhead scaling remains exponential in the number of Pauli noise operators $|\mathcal{K}|$ within the sparse model assumption. In this sense, our scheme does not circumvent the need for advancing quantum error correction.\\

\noindent \textbf{Quantum circuit implementation of NAFQA.}~ The quantum circuit for NAFQA is built iteratively: the parameters for the $s+1$-th layer are determined from measurement on a quantum circuit built up to the $s$-th layer. Each layer contains the unitaries $U_p$, $U_d(\beta_s)$, and the noise channel $\mathcal{F}_s = \Lambda_s \circ \Lambda'_s$, as shown in Fig.~\ref{Fig: Schematic_plot} (c). For the next layer, we define the control field $\beta_{s+1} = - A_{s}$ and the modified noisy map $\mathcal{F}_{s+1} (\cdot) = \Lambda \circ \Lambda' (\{\Gamma(s+1)\})$ that contains error probabilities $\Gamma(s+1)_k = - \{(C(s)_k\}_{k\in \mathcal{K}}$, where the mathematical form of the noise channel $\Lambda'(\cdot)$ is given in Methods. The parameters $C(s)$ and $\Gamma(s+1)$ are used for the Trotter steps $s$ and $s+1$, respectively. The method to simulate the decay rates $\Gamma$ in an ideal quantum circuit at the Trotter step $s$ is summarized in Algorithm \ref{algor_pseudo}, and is further discussed in Methods. We repeat the process for a finite number of Trotter steps $s$ and find that this feedback-based procedure ensures a monotonic decrease in energy with additional Lyapunov controls, where the noise channels drive the system faster toward the virtual ground state.

% \begin{figure}[t!]
%     \centering
%     \includegraphics[scale=0.31]{final_plots/large_s.pdf} 
%     \caption{\textbf{Numerical results for the 5-qubit Maxcut problem.}
%     }\label{large_s}
% \end{figure}

% To kick-start the feedback process, we initialize the first layer of the quantum circuit by incorporating the inherent noise in the quantum circuit. Using randomized benchmarking~\cite{EmersonPRA2016, Blatt_nat_comm_2019}, we modeled the noise as stochastic Pauli noise channels and characterized its error probabilities with the cycle error reconstruction method~\cite{Flammia2020}. Here, we have assumed that the sparse noise model given in Eq.~\eqref{noise_channel_Lambda1} is the same for each layer, which leads to $\Lambda = \Lambda_1 = \Lambda_2 = \cdots = \Lambda_s$. This assumption is justified since the unitaries corresponding to $H_p$ are fixed and the unitary corresponding to $H_d$ contains only single-qubit terms. However, to accurately implement noise, one must learn the noise models $\Lambda_s$ at each layer.

To kick-start the feedback process, we initialize the first layer of the quantum circuit by incorporating the inherent noise in the quantum circuit as stochastic Pauli noise channels. Here, we have assumed that the Pauli sparse noise model given in Eq.~\eqref{noise_channel_Lambda1} is the same for each layer, which leads to $\Lambda = \Lambda_1 = \Lambda_2 = \cdots = \Lambda_s$. This assumption is justified since the unitary corresponding to $H_p$ is fixed and the unitary corresponding to $H_d$ contains only single-qubit terms. However, to accurately implement the intrinsic noise, one must learn the noise models $\Lambda_s$ at each layer.

A key feature of our NAFQA protocol is that the coefficients $\Gamma_k(t)$ can take negative values. While such values correspond to unphysical noisy channels, they nonetheless satisfy the Lyapunov condition in the OQS framework. This type of OQS dynamics with negative $\Gamma$ can be implemented, as described in Algorithm \ref{algor_pseudo}.

\begin{figure*}[t!]
    \begin{tabular}{c c c}
    \includegraphics[scale=0.38]{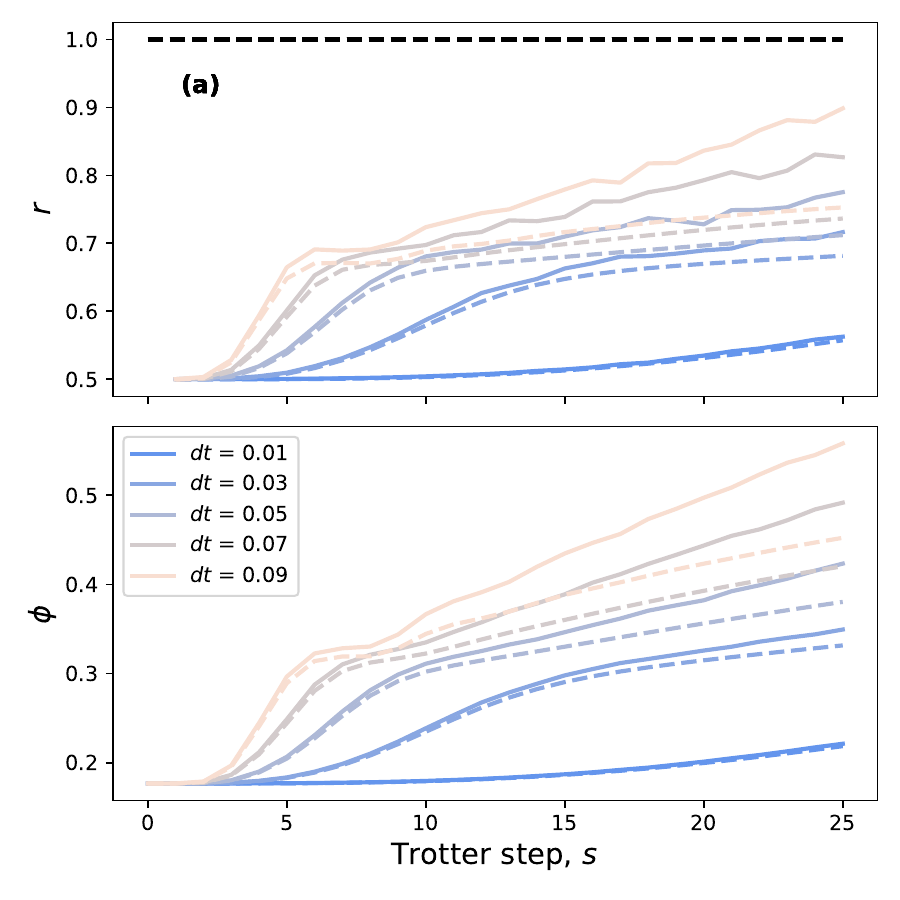} &
    \includegraphics[scale=0.38]{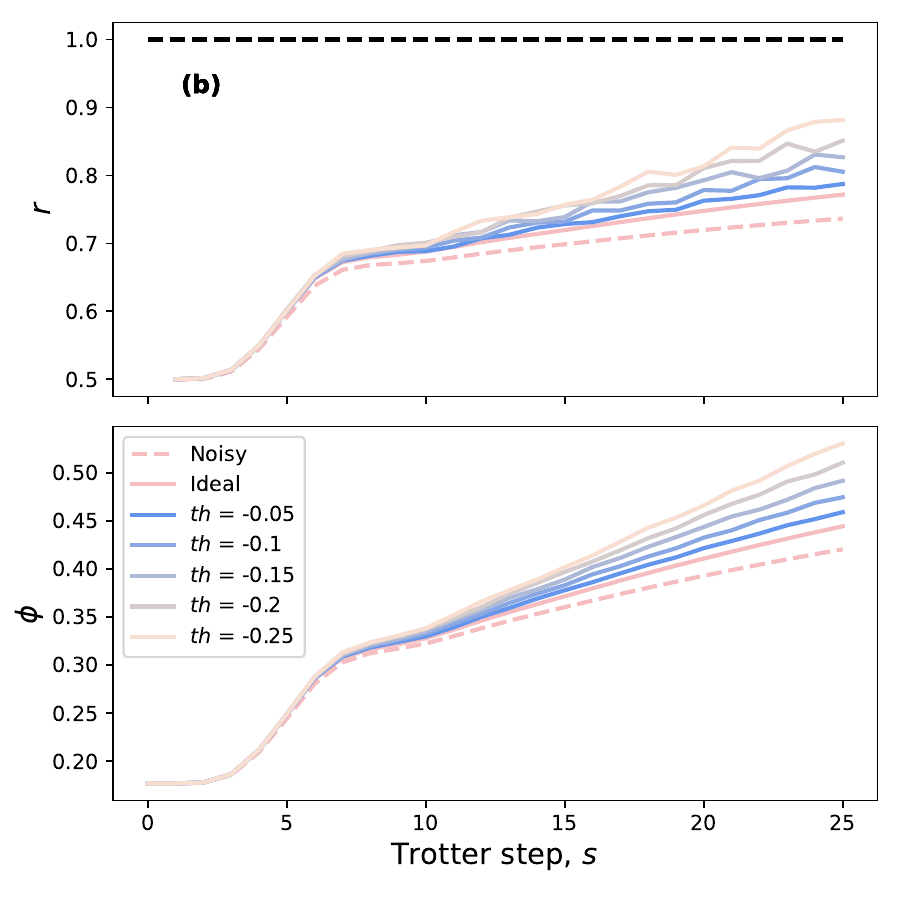} &
    \includegraphics[scale=0.38] {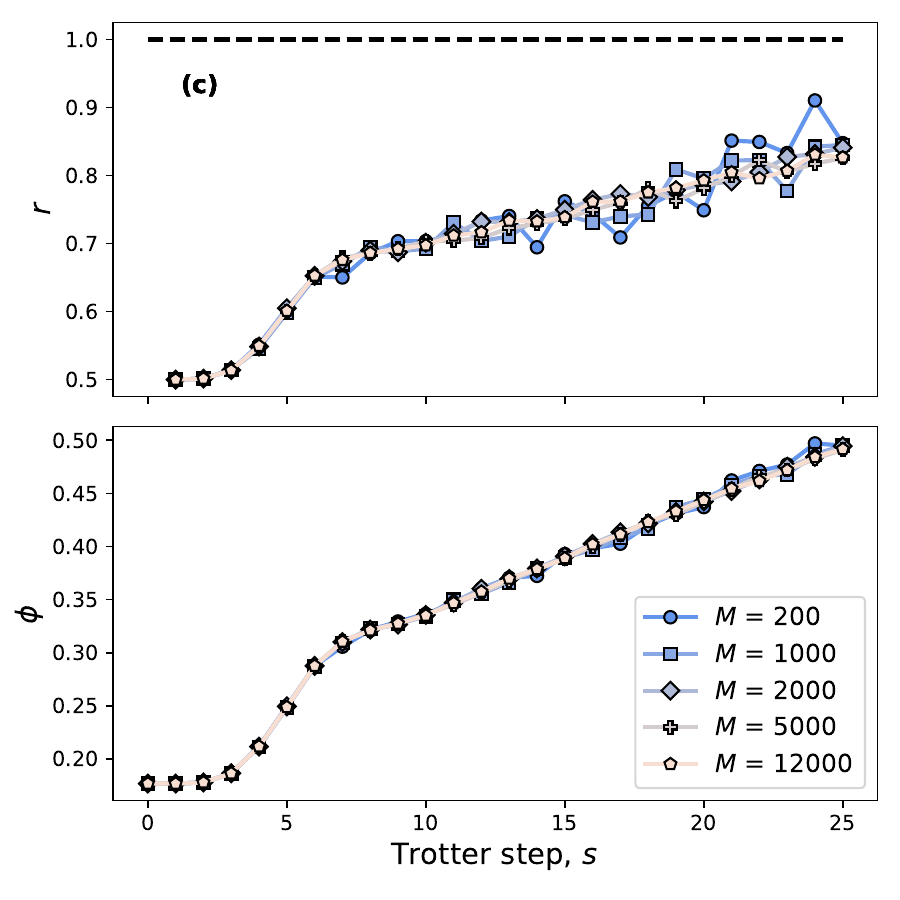}
    \end{tabular}
    \caption{\textbf{Time evolution of approximate ratio $r$ and success probability $\phi$ with  different parameters involved in the NAFQA protocol.} We considered several cases: (a) varying the time steps $dt$ with a fixed sample size, $M = 12000$, (b) different threshold values of $\Gamma(t)$ from $\text{th} = -0.05$ to $\text{th} = -0.25$ with $dt = 0.07$ and $M = 12000$. The dotted and solid lines indicate the noisy and NAFQA evolutions, respectively. (c) Various numbers of samples $M$ are used for a fixed time step $dt = 0.07$ and threshold value of $\text{th} = -0.15$ .}\label{Fig: panel plot}
\end{figure*}

\begin{algorithm}[H]
\caption{Quantum circuit simulation for non-Markovian dynamics}
\KwIn{Initial state $|\psi_m^{(s-1)}\rangle$, noiseless ideal unitary $U_s$ at the $s^{\text{th}}$ Trotter layer, number of samples $M$, Trotter-step $dt$}
% \KwOut{Final state $\rho_f$}

\For{$0\leq m \leq M$}{
        {1. Compute  $\ket{\psi_m}^{(s)} = U^{(s)} \ket{\psi_m}^{(s-1)}$ \\

        2. Sample $P_k$ with probability $r_k dt$, with $r_k = {|\Gamma_k|}$  \\
        Compute  $\ket{\psi_k}^{(s)} = \sqrt{|\Gamma_k|}P_k \ket{\psi_k}^{(s)}/ {\sqrt{r_k}}$ \\
        Update sign: $s^{(s)}_m \gets s^{(s-1)}_m \cdot \frac{\Gamma_k}{|\Gamma_k|}$\\
        
        \textbf{Or}\\

        2. Sample $I$ with probability $r_{\text{res}} = 1-\sum_k r_k dt $ \\
        Compute  $\ket{\psi_k}^{(s)} = \ket{\psi_k}^{(s)} / {\sqrt{r_{\text{res}}}}$\\
        Update sign: $s^{(s)}_m \gets s^{(s-1)}_m (+1)$}\\

    3. Compute: $\rho^{(s)}_m = s^{(s)}_m \cdot |\psi_m^{(s)}\rangle \langle \psi_m^{(s)}|$ , $\mathcal{N}^{(s)} = \sum_m^{{M}} s_m^{(s)} \braket{\psi_m^{(s)} | \psi_m^{(s)}}$ 
}
\Return $\rho^{(s)}_f = \frac{1}{\mathcal{N}^{(s)}} \sum_{m=1}^{M} \rho^{(s)}_m$ \label{algor_pseudo}
\end{algorithm}

% Thus, on average, this channel is completely positive and trace-preserving.

\noindent \textit{Protocol for calculating the expectation values with negative decay rates in experiments.}~{The idea behind QPD is to represent an ideal circuit as a quasiprobabilistic mixture of noisy circuits~\cite{Gambetta2017PRL}. With this, one can efficiently measure the bias-free expectation value with correct classical post-processing of the measurement outcomes. This idea can be extended to decompose any arbitrary physical or nonphysical map $\Lambda'_s(\rho)$. It can be expressed as a linear combination of noisy operations $\{\mathcal{\tilde{U}}\}$ that can be implemented on quantum hardware, 
\begin{equation}
    \Lambda'_s(\rho) = \sum_i \eta_{s,i} \mathcal{\tilde{U}}_i(\rho) \approx \mathcal{N} _s \sum_i \frac{|\eta_{s,i}|}{\mathcal{N} _s} \text{sgn} ~ (\eta_{s,i}) \mathcal{\tilde{U}}_i,\label{samp_over}
\end{equation}
where $\eta_{s,i}$ can be negative, $\text{sgn}(\cdot)$ denotes the sign function, and $\mathcal{N} _s = \sum_i |\eta_{s,i}|$ is the sampling overhead that satisfies $\mathcal{N}_s \geq 1$. The probability $p_{s, i} = \frac{|\eta_{s,i}|}{\mathcal{N}_s}$ is non-negative and sums up to 1. Given an input state $\rho$, the expectation value of an operator $O$ after the desired channel $\Lambda'_s(\rho)$ can be written as
\begin{equation}
    \braket{O}_{\text{eff}} = \text{tr}[O\Lambda'_s(\rho)] = \mathcal{N}_s \sum_i p_{s,i} ~ \text{sgn} (\eta_{s,i}) \text{tr}[O~\mathcal{\tilde{U}}_i(\rho)]. \label{qpd_eqn}
\end{equation}

The QPD is implemented using the following procedure. For each channel $\Lambda'_s(\rho)$, the noisy operation $\mathcal{\tilde{U}}_i$ is sampled with probability $p_{s,i}$, and the measured expectation value is multiplied by the parity, $\text{sgn}(\eta_{s,i})$. After sampling over multiple estimators, the computed unbiased expectation value $\braket{O}_{\text{eff}}$ is obtained after multiplying with the overhead coefficient $\mathcal{N}_s$, as given in Eq.~\eqref{qpd_eqn}. Note that the measurement $\text{tr}[O~\mathcal{\tilde{U}}_i(\rho)]$ taken after each sample is executed on a quantum computer and the final expectation value of the observable $\braket{O}_{\text{eff}}$ is calculated on a classical computer after scaling with the signs and the sampling overhead. However, the computation of the expectation value comes with an additional overhead as the variance of the estimator increases by a factor of $\mathcal{O}(\mathcal{N} ^2)$~\cite{Jinzhao_Sun_PhysRevApplied_2021}. The total overhead for $L$-layers becomes $\mathcal{N}_{\text{tot}} = \prod_{s = 1}^{L}\mathcal{N} _s$. QPD has a runtime that scales roughly as $(1 + \frac{15}{8} \epsilon)^{NL}$, where $N$ is the number of qubits, $L$ is the circuit depth, and $\epsilon$ is the two-qubit depolarizing error~\cite{vandenBerg2023_natphys}. This exponential scaling does not mean that these protocols are without value; in fact, lowering hardware error rates $\epsilon$ can provide a promising pathway to quantum advantage~\cite{aharonov2025importanceerrormitigationquantum, zimborás2025mythsquantumcomputationfault}. Rapid developments in reducing variance and run-time cost using tensor network error mitigation~\cite{fischer2024dynamicalsimulationsmanybodyquantum} and
shaded light cones~\cite{eddins2024lightconeshadingclassicallyaccelerated} are also demonstrated, opening new doors for the introduction of algorithms incorporating these techniques.} 

Although we model noise using a general stochastic Pauli channel, we emphasize that our approach is broadly applicable and can be implemented with any noise model. The only requirement is that the noise that affects each layer must be known or characterized in advance. A more detailed discussion of noise models beyond stochastic Pauli channels, particularly those involving amplitude and phase damping, is presented later in the paper.\\

\noindent \textbf{Optimization problem.}~We demonstrate the effectiveness of our NAFQA algorithm by applying it to two illustrative examples, the Maxcut problem and the transverse field Ising models. The initial state $\ket{\psi_0}$ is prepared in the state $\ket{+}^{\otimes N}$ for a $N$ qubit system. The performance of our algorithm is measured based on two figures of merit: the approximation ratio, $r = {\braket{H_p}}/{\braket{H_p}_{\text{min}}}$ and the success probability, $\phi = \bra{\psi_{\text{gs}}} \rho \ket{\psi_{\text{gs}}}$, where $\rho$ is the density matrix and $\ket{\psi_{\text{gs}}}$ is the ground state obtained by diagonalizing the problem Hamiltonian $H_p$. Both figures of merit $r$ and $\phi$ vary from 0 to 1, with $r = \phi \approx 1$ corresponding to the approximate solution of the ground state.

As a first example, we consider the unweighted Maxcut problem, where the Hamiltonian for $N$ qubits is given by $H_p = -\sum_{i,j\in \mathcal{E}} \frac{1}{2}(1 - Z_iZ_j)$. The control Hamiltonian is chosen as $H_d = \sum_{j = 1}^{N} X_j$. Given $H_p$ and $H_d$, the commutator in Eq.~\eqref{closed_system} takes the form $i[H_d,H_p] = \sum_{i,j\in \mathcal{E}} Y_iZ_j + Z_i Y_j$. Here, $X_j$, $Y_j$, and $Z_j$ are the Pauli operators that act on the qubit $j$. 

\begin{figure}[t!]
\includegraphics[scale=0.24]{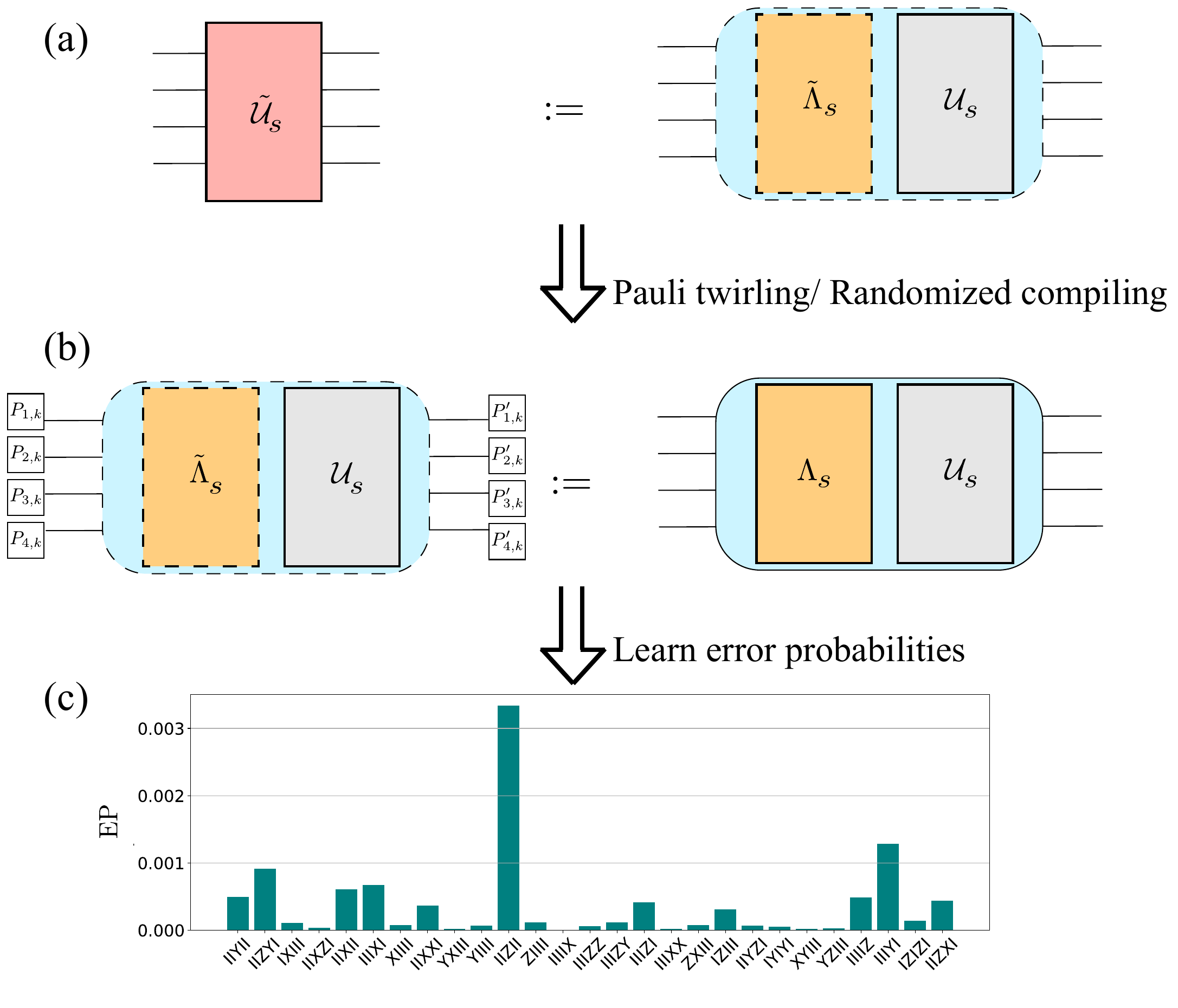} 
    \caption{\textbf{Overview of the noise modeling and learning.} (a) Ideal unitary at the $s^{\text{th}}$ layer whose action is denoted by $\mathcal{U}_s$ and its implementation on NISQ is $\tilde{\mathcal{U}}_s$, decomposed as $\tilde{\mathcal{U}}_s = \mathcal{U}_s \circ \tilde{\Lambda}_s$. (b) Pauli twirling or randomized compiling shapes the noise channel $\tilde{\Lambda}_s$ to an effective stochastic Pauli channel ${\Lambda}_s$ by randomly applying one-qubit Paulis before and after the noisy layer. For Clifford operators $U$, we obtain $ P'_{n,s} = UP_{n,s}U^{\dagger}$, where $n$ is the $n^{\text{th}}$ qubit. (c) Learned error probabilities of the 5-qubit emulated IBM-Vigo simulator.
    } \label{Fig: twilring_plot}
\end{figure}

In Fig.~\ref{one_decay_op}, we present the results of the NAFQA algorithm and compare them with simulations of the FQA designed for closed-system dynamics, both in the presence and absence of noise. ``Ideal" refers to noiseless FQA simulations (Fig.~\ref{Fig: Schematic_plot} (a)), while ``Noisy" simulations correspond to the FQA simulations with the stochastic Pauli noise channels (Fig.~\ref{Fig: Schematic_plot} (b)). Noisy simulations are equivalent to solving Eq.~\eqref{rho_evolution} with the learned error probabilities $\{\lambda\}$ of quantum hardware, while still enforcing the Lyapunov condition defined for an isolated system, as introduced in Eq.~\eqref{closed_system}. We learned the error probabilities $\{\lambda\}$ of the 5-qubit IBMVigo simulator using Qiskit~\cite{Qiskit2021} and plotted them in Fig.~\ref{Fig: twilring_plot} (c). The schematic circuit diagrams for FQA and NAFQA in the NISQ devices are illustrated in Figs.~\ref{Fig: Schematic_plot} (b) and (c), respectively. 

% We emphasize that Trotterized implementations of ideal FQAs on quantum hardware have not consistently shown a monotonic decrease in energy with Trotter steps~\cite{SarovarPRL2022, RKMalla_PRR_2024}. This inconsistency arises due to the inherent susceptibility of quantum systems to noise and interactions with their surrounding environment~\cite{Breuer_book}. as illustrated in Figs.~\ref{one_decay_op} (a) and (c)

We begin by characterizing the approximation ratio, $r$, and the success probability, $\phi$, as a function of $s$. We observed that with NAFQA, $r$ ($\phi$) increases from $0.5 ~ (0.17)$ at $s = 0$ to $0.97$ ($0.64$) at $s = 41$. In contrast, for the noisy (ideal) simulations, we obtain $r \approx 0.77 (0.82) $ and $\phi \approx 0.48 (0.53)$ at $s = 41$ (Fig.~\ref{one_decay_op} (a) and (c)). This corresponds to an improvement of $26\%$ for $r$ and an improvement of $33\%$ for $\phi$ with our methods compared to the noisy FQA simulations. For the NAFQA simulations, here we modify only one Pauli term as a function of $s$, and set the error probabilities of all the other terms to zero. This shows that even a single modified error probability can accelerate the system's evolution towards the virtual ground state, highlighting the effectiveness and adaptability of our algorithm. Figure~\ref{one_decay_op} (b) shows the time dependence of the control parameters $\beta$ and $\Gamma$. Enforcing QLC in NAFQA yields negative $\Gamma$ rates, which are obtained by the non-Markovian Monte Carlo solver in Qutip~\cite{JOHANSSON_qutip_2013}. 

The noisy simulations demonstrate non-monotonic behaviour in both $r$ and $\phi$ (Fig.~\ref{one_decay_op} (e)), and it decays to $r ~ (\phi) \approx 0.65 (0.35)$ after $s = 500$ steps. This is expected as they do not satisfy the QLC condition, see Eq.~\eqref{lyapunov_lambda}. The convergence of NAFQA with respect to the number of samples $M$ grows exponentially with time, as shown in Eq.~\eqref{error_mitigation_cost}. To quantify this behavior, we define the relative error $\delta$ as the deviation from the normalized trace of $1$ as, $\delta = (\text{Tr} -1 ) \times 100$, where $\text{Tr}$ is the trace of the state. Figure~\ref{one_decay_op} (d) shows that $\delta$ increases with $s$ for a fixed number of samples, and decreases in a particular $s$ as $M$ increases. Next, we plot the time-averaged $\delta$ as a function of $M$ on a double logarithmic scale, and it scales approximately as $\delta \propto 1/\sqrt{M}$. This verifies that running NAFQA at long timescales requires an exponentially large number of samples (Eq.~\eqref{error_mitigation_cost}). Therefore, it is not possible to show a convergence of $r$ and $\phi$ as a function of $s$.

The convergence of NAFQA depends on the magnitude of the negative decay rates as well (Eq.~\eqref{error_mitigation_cost}). To regulate the decay rates $\Gamma_k$, we establish a threshold of $-0.15$ as demonstrated in Fig.~\ref{one_decay_op} (b) i.e., if $\left(\braket{ P_k^\dagger H_p P_k } - \braket{H_p}\right) > 0.15$, we set $\Gamma_k = - 0.15$; otherwise, we assign $\Gamma_k (t)= - \left(\braket{ P_k^\dagger H_p P_k } - \braket{H_p}\right)_t$. We note that the choice of $-0.15$ as the threshold is arbitrary. However, a finite threshold value mitigates the need for a prohibitively large number of trajectories, which would otherwise increase the computational cost as $\Gamma_k$ decreases. 

We now proceed to examine the performance of NAFQA and compare it with noisy evolutions through a series of numerical simulations. Figure~\ref{Fig: panel plot} illustrates the behavior of $r$ and $\phi$ for different values of time steps $dt$, threshold value `$\text{th}$', and sample sizes $M$. Increasing the Trotter step $dt$ improves $\phi$ and $r$ monotonically as a function of $s$, as shown in Fig.~\ref{Fig: panel plot} (a). Decreasing the threshold value similarly increases the performance of the protocol, as seen in Fig.~\ref{Fig: panel plot} (b). Finally, Fig.~\ref{Fig: panel plot} (c) reveals that both $r$ and $\phi$ converge with increasing $M$, indicating that statistical errors do not accumulate and the protocol remains robust in the large-sample limit. 

We also demonstrate the performance of FQA in the ideal and noisy framework by varying the control field $\beta(t)$ and the Trotter-step $dt$, as illustrated in Fig.~\ref{beta_change_purity}. We find that making $\beta(t)$ more negative or increasing $dt$ speeds up evolution ($r$ and $\phi$ increase) for short times. However, there is a bound for both $\beta$ and $dt$, and one should not cross these limits, otherwise the parameters start to oscillate very frequently and the FQA protocol does not work. Although we provide bounds on the parameters (see SI), tighter bounds would be desirable to optimize the choice of $\beta$ and $dt$, which could be investigated in future work. We note that simply making $\beta(t)$ more negative or increasing $dt$ does not solve the heating problem. 
Our results indicate that FQAs implemented on NISQ devices will inevitably evolve into mixed states and will not converge to their pure ground states.
This is verified by the monotonic decrease of the purity $\mathcal{P}(t) = \text{Tr}[\rho(t)^2]$, as shown in the insets in Fig. \ref{beta_change_purity}. Thus, our NAFQA protocol provides a practical approach to finding ground-state properties on NISQ devices.\\

\begin{figure}[t!]
    \centering
    \includegraphics[scale=0.425]{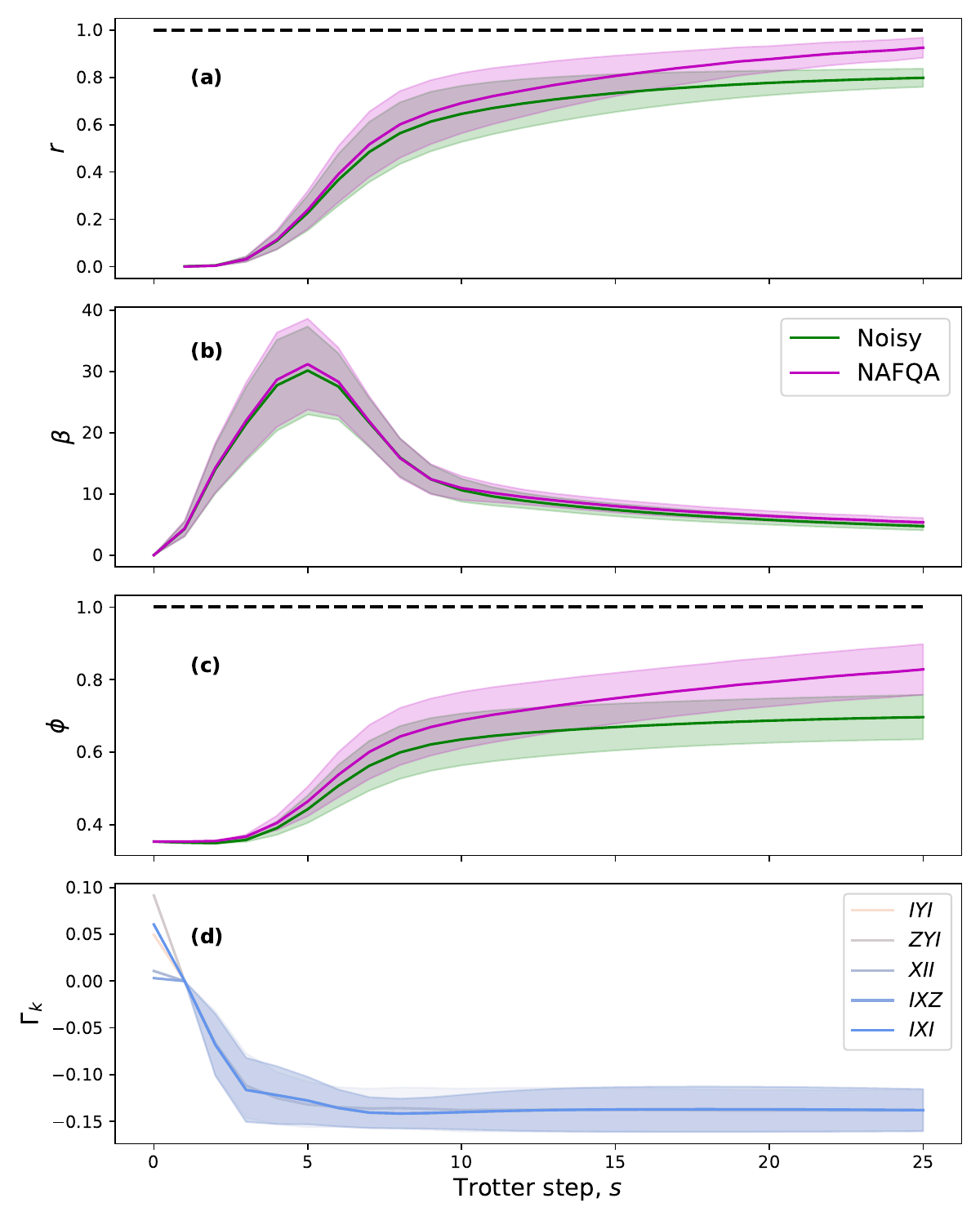} 
    \caption{The performance of NAFQA against the noisy FQA with error probabilities given in Fig.~\ref{second_err_probs}. The mean and standard error (a) approximation ratio $r$, (b) control field $\beta$, and (c) success probability $\phi$ for the 3-qubit spin glass Hamiltonians over 25 instances are illustrated. The numerical simulations are carried out with a time step of $dt = 0.005$, yielding NAFQA results from 5000 trajectories. (d) error probabilities $\Gamma_k(t)$ for $k\in{IYI, ZYI, XII, IXZ, IXI}$. }\label{TFIM_plot}
\end{figure}

\noindent \textbf{Spin glass Hamiltonians.}~We now present the performance of our algorithm for spin-glass Hamiltonians over several instances. We consider an all-to-all spin-glass Hamiltonian that is expressed as $H_p = \sum_{m<n}J_{mn}Z_m Z_n + \sum_l h_l Z_l$, where the control Hamiltonian $H_d$ takes the form of $H_d = - \sum_{j = 1}^{3} X_j$. With these choices of $H_p$ and $H_d$, the commutator becomes $i[H_d,H_p] =  -2 \left(\sum_{i<j}J_{ij} (Y_i Z_j + Z_i Y_j) + \sum_{i} h_i Y_i \right)$. The resulting mean and standard error of $r$, $\phi$, and $\beta$ are shown in Fig.~\ref{TFIM_plot}. The numerical simulations are performed over 25 instances for 5000 trajectories. The coupling constants $J_{mn}$ and the transverse field strengths $h_l$ are randomly chosen from a uniform distribution of $[-1,+1]$. Similarly to the Maxcut problem, NAFQA successfully achieves a faster convergence to the ground state of spin-glass problems. Noisy simulations are performed with the weights given in Fig.~\ref{second_err_probs}. However, here we have considered five additional Lyapunov controls that decide the modified error probabilities $\Gamma_k$ ($k = 1,2,3,4,5$) compared to only one for the Maxcut problem. As shown in Fig.~\ref{TFIM_plot}, a threshold of $-0.15$ is taken for all decay rates. We note that $\beta(t)$ varies smoothly throughout the simulations and that the modified rates $\Gamma_k(t)$ are the key factors that enable the extraction of properties of the ground-state on shorter timescales.\\

\noindent \textbf{\large Discussion and Outlook}\\
\noindent This work opens several promising avenues for future exploration. For instance, the control protocol for OQS could be designed to steer the system toward a decoherence-free subspace~\cite{Wang2010}; see SI. Our approach can also be explored beyond the stochastic noise channels, especially in the presence of both unital and non-unital noise or when the noise is integrated with quantum gates~\cite{Bassi_prr_2023}. This work may also have applications in fast charging and discharging of quantum batteries~\cite{Shastri2025} and in rapid thermalization of quantum states with shortcuts to adiabaticity~\cite{Alipour2020shortcutsto,Yin2022}. Furthermore, recent advances in quantum algorithms for cooling and dissipative many-body ground-state preparation~\cite{lin_lin_review_2025, barbara_cooling} may be relevant and complementary to our approach, offering a tantalizing prospect for their combination. \\

\noindent \textbf{\large Conclusion}\\
\noindent Because quantum systems are inherently open, it is both natural and advantageous to design control protocols and quantum algorithms grounded in OQS theory. At the same time, intrinsic device noise remains a major obstacle to quantum simulation. A crucial first step toward overcoming this challenge is to characterize the intrinsic noise relevant to a given algorithm, using techniques such as randomized benchmarking and error reconstruction. As environmental noise cannot be completely eliminated, developing algorithms that integrate and exploit the intrinsic noise of NISQ devices is highly desirable. In this work, we have established a noise-assisted quantum optimization framework and demonstrated its effectiveness on emulated IBM quantum processors. By combining a sparse learning protocol with the pseudo-Lindblad quantum trajectories approach, we simulated non-Markovian, time-dependent open quantum systems (OQS). Applying our algorithm to the MaxCut and spin-glass Hamiltonians, we achieved population transfer toward the virtual ground state. The simulation of negative rates faces an exponential overhead in the system size and requires extensive sampling and measurement statistics for efficient implementation. Future efforts to mitigate these limitations may form the basis for realizing quantum advantage.\\

\noindent \textbf{\large Methods}\label{methods}\\
\noindent \textbf{Learning the Pauli-Lindblad noise model.}\label{Pauli_noise_learning}~Here, we provide an overview of the noise characterization process corresponding to the noise associated with each layer of a digital quantum circuit. The unitary evolution of quantum gates on current NISQ devices is hindered by noise and dissipation. The application of an ideal unitary $U$ on qubits makes the evolution $\mathcal{U}(\rho) = U\rho U^{\dagger}$. Since quantum gates are noisy, a noisy channel can be defined as the composition of the ideal unitary $U$ and noisy operation $\tilde{\Lambda}$, that is, $\mathcal{\tilde{U}} = U \circ \tilde{\Lambda}$. The Pauli twirling simplifies the noise model $\tilde{\Lambda}$, and converts any arbitrary noise on the quantum hardware into stochastic Pauli channels~\cite{Flammia2020}, as illustrated in Fig.~\ref{Fig: twilring_plot}. It is obtained by sampling random instances of Pauli operators around the noisy layers. Algebraically, this is expressed as
\begin{equation}
    \Lambda(\cdot) = \mathbb{E}_i \left[ P_i^\dagger \tilde{\Lambda} \left( P_i \cdot P_i^\dagger \right) P_i \right].\label{noise_map}
\end{equation}
For the dissipator considered in Eq.~\eqref{pauli_channel}, it is shown that the sparse Pauli Lindblad noise channel can be expressed as~\cite{vandenBerg2023_natphys}
\begin{equation}
    \Lambda(\rho) = \prod_{k\in\mathcal{K}} \left( \omega_k \cdot + (1-\omega_k) P_k \cdot P_k^{\dagger} \right) \rho, \label{noise_channel_Lambda}
\end{equation}
where $\omega_k = (1+e^{-2\lambda_k})/2$ and the error probabilities $\{\lambda\}$ characterize the intrinsic noise of the quantum hardware. For a detailed exposition of the sparse Pauli learning process, we refer the reader to~\cite{vandenBerg2023_natphys, Ferracin2024efficiently}.\\

\noindent \textbf{Pseudo-Lindblad quantum trajectory unraveling.}\label{PLQT}~Here, we present the PLQT approach~\cite{Eckardt_PRL} and discuss how it can be implemented in a digital quantum circuit to simulate non-Markovian dynamics. 

The simulation of a GKSL master equation requires the evolution of a density matrix $\rho$ of size $d_H^2$ when the Hilbert space has dimension $d_H$. However, simulating a state (wavefunction) instead of $\rho$ is numerically more efficient, since the state has dimension $d_H$. Therefore, the MCWF method was proposed as an alternative approach to simulate a master equation, taking an ensemble average over multiple trajectories~\cite{Dalibard_etal_prl, Carmichael1993Open}. After a time step $dt$, the state either evolves under the effective Hamiltonian $H_{\text{eff}}$ as
\begin{equation}
    \ket{\psi(t + dt )} \propto (1-i dt H_{\text{eff}}) \ket{\psi(t)} \label{det_evol},
\end{equation}
with probability $1 - \sum_k r_k(t) dt $, or a quantum jump occurs with probability $r_k(t) = || L_k \ket{\psi(t)} || ^ 2 / || \ket{\psi(t)} || ^ 2$ and the state evolves as
\begin{equation}
    \ket{\psi(t + dt )} \propto L_k \ket{\psi(t)}.\label{jump_evol}
\end{equation}
Here, the effective non-Hermitian Hamiltonian is $H_{\text{eff}} = H - \frac{i}{2} \sum_k L_k^{\dagger} L_k$. The stochastic Schr\"odinger equation for this unraveling can be expressed in the It\^{o} formalism as  \begin{equation}
    \ket{d\psi} = \sum_k dN_k \left( \frac{~~L_k}{\sqrt{r_k(t)}} \ket{\psi} - \ket{\psi} \right) + dt \left( \sum_k \frac{r_k(t)}{2} - i  H_{\text{eff}}\right) \ket{\psi}.
\end{equation}
$dN_k$ represents independent jump processes, with $dN_k$ being one (zero) in the event of a quantum jump (no quantum jump). 

The GKSL equation states that if $\mathcal{L}$ is the Liouvillian governing the OQS evolution according to the master equation $d\rho / dt = \mathcal{L} (\rho) $, then there exists a CPTP quantum channel of the form $\mathcal{E} (\rho) = \text{exp}(\mathcal{L}t)$ for the evolution of $\rho$. The noisy gates in a quantum circuit can be simulated with this quantum trajectory approach by writing a CPTP channel with the Kraus representation as 
\begin{equation}
    \mathcal{E} (\rho) = \sum_k M_k (dt) \rho M_k ^{\dagger} (dt),
\end{equation}
where $\sum_k M_k(dt) M_k ^{\dagger} (dt) = I$. This map can be implemented in a quantum circuit as a stochastic map, where the state is updated to $\ket{\psi'} = \frac{1}{\sqrt{r_k}} M_k(dt) \ket{\psi}$ with probability $r_k = |\braket{{\psi} | M_k ^{\dagger}(dt) M_k(dt) | {\psi}}|^2$. The Kraus operators representing Eqs.~\eqref{det_evol} and \eqref{jump_evol} for a Trotter step of $d t $ are given by 
\begin{equation}
    M_0(dt) = I -i H_{\text{eff}} dt ~ , ~ M_k(dt) = \sqrt{dt} L_k~, ~ k\neq0.
\end{equation}

% Utilizing this noise map,
We extend this formalism to simulate layer-dependent non-Markovian dynamics on a digital quantum computer with the PLQT approach. It relies on extending the Hilbert space of the system by an additional classical bit $s \in \{-1, +1\}$. Since we are interested in the negative decay rates in the GKSL equation, Eq.~\eqref{na--qlc} can be modified as 
\begin{equation}
    \frac{d \rho}{dt} = -i [H, \rho] + \sum_k s_k (L_k \rho L_k^\dagger - \frac{1}{2} L_k^\dagger L_k \rho - \frac{1}{2} \rho L_k^\dagger L_k ),
\end{equation}
where each Lindblad operator $L_k$ is associated with a stochastic Pauli channel. This involves a positive relaxation strength $|\Gamma_k|$, as $L_k = \sqrt{|\Gamma_k|} P_k$. The signs $s_k = \Gamma_k / |\Gamma_k| = \pm 1$ are multiplied in the classical post-processing step to simulate the non-Markovian dynamics. In the It\^{o} formalism, the stochastic Schr\"odinger equation is described as 
\begin{equation}
    \ket{d\psi} = \sum_k dN_k \left( \frac{~s_k L_k}{\sqrt{r_k(t)}} \ket{\psi} - \ket{\psi} \right) + dt \left( \sum_k \frac{r_k(t)}{2} - i  H_{\text{eff}}'\right) \ket{\psi}, 
\end{equation}
where $H_{\text{eff}}' = H - \frac{i}{2} \sum_k s_k L_k^{\dagger} L_k$. As we are only concerned with modifying the noise channel, we do not consider the contribution of the unitary part that comes from the Hamiltonian $H$. Furthermore, for the Pauli noise channels considered in Eq.~\eqref{pauli_channel}, we have $\sum_k L_k^\dagger L_k \propto I$. Therefore, the non-Hermitian effective Hamiltonian becomes $H_{\text{eff}} = - \frac{i}{2} \sum_k s_k |\Gamma_k| I $, which introduces a global phase factor and can be ignored. The state undergoes a quantum jump 
\begin{equation}
\begin{split}
    \ket{\psi^{(k)}(t + dt)} & = \frac{\sqrt{|\Gamma_k|}P_k \ket{\psi(t)}}{\sqrt{r_k(t)}}, \\
    s^{(k)}(t + dt) & = \frac{\Gamma_k}{|\Gamma_k|} s^{(k)}(t),
\end{split}
\end{equation}
with probability $r_k(t) dt$ and the identity operator $I$ acts with the remaining probability. The final-state density operator after evolution is obtained as $\rho(t) = \overline{s(t)\ket{\psi(t)}\bra{\psi(t)}}$. For the choice of Lindblad operators, we have $r_k = {|\Gamma_k|}$, which cancels out in the ensemble average. For a finite sample size $M$, the state at time $t$ can be renormalized as $\rho_{{M}}(t) = \frac{1}{\mathcal{N}} \sum_m^{{M}} s_m \ket{\psi_m(t) } \bra{\psi_m(t) }$, where $\mathcal{N} = \sum_k^{{M}} s_m \braket{\psi_m(t) | \psi_m(t)}$. Here, the Kraus operators can be written as a mixed unitary channel with $M_0 \propto I$ and $M_k \propto P_k$. \\

\noindent \textbf{Quasiprobability decomposition for the experimental implementation.}~The noise map $\Lambda'(\rho)$ corresponding to the Liouvillian $\mathcal{L}(\rho) = \sum_{k\in \mathcal{K}} s_k |\nu_k| (P_k \rho P_k^{\dagger} - \rho)$  can be simplified to~\cite{vandenBerg2023_natphys} 
\begin{equation}
    \Lambda'(\rho) = \prod_{k\in\mathcal{K}} \left( \omega'_k \cdot + s_k (1-\omega'_k) P_k \cdot P_k^{\dagger} \right) \rho,
\end{equation}
where $\omega'_k = (1+e^{-2 |\nu_k| })/2$, and the sampling overhead $\mathcal{N} = \omega'_k +  s_k (1-\omega'_k)$  is consistent with Eq.~\eqref{samp_over}. The sampling algorithm to virtually implement the $\Lambda'$ is as follows:
\begin{enumerate}
    \item For every $k\in\mathcal{K}$, sample the identity matrix with probability $\omega'_k$, and the Pauli matrix $P_k$ with probability $1-\omega'_k$.
    \item Store the sign $s_k$ when the Pauli matrix is sampled. 
    \item Measure the expectation value for each sampling at the end of the circuit (executed on a quantum computer) and multiply its sign (on a classical computer).
    \item Measure the final expectation value after sampling over all the estimators and finally scale it with the sampling overhead $\mathcal{N}$ (on a classical computer), as given in Eq.~\eqref{qpd_eqn}.
\end{enumerate}

The targeted coefficients $\Gamma_k(t)$ in Eq.~\eqref{na--qlc} for each layer can be obtained by integrating the noise learning process with the QPD approach. Utlizing the property of channel operations, we have $\Gamma_k(t) = \lambda_k + \nu_k(t)$. Using $\{\nu_k(t)\}$ and $\{\lambda_k\}$ the modified stochastic noisy map $\mathcal{F}_s$ can be expressed as $\mathcal{F}_s = \Lambda_s \circ \Lambda_s'$, as given in Eq.~\eqref{non_Markovian-ckt}. Here, the subscript $s$ is used for the Trotter step $s$. The state of the system at time $t+dt$ is then given by 
\begin{equation}
    \rho(t+ dt) = \mathcal{\tilde{U}}_s \mathcal {F}_s (\rho(t)),
\end{equation}
where $\{\mathcal{\tilde{U}}\}$ are the set of implementable noisy operations. Starting from an initial pure state $\rho_0$ at time $t = 0$, the final state after performing NAFQA for $s$ Trotter steps will be closer to the approximate virtual ground state of the system, i.e., $\rho(s ~dt) \approx \rho_{\text{ground state}}$.\\

% \noindent \textbf{Competing interests}\\
% The authors declare no competing interests.
% \vspace{0.2cm} 

% \noindent \textbf{Code availability}\\
% Code available upon reasonable request.
% \vspace{0.2cm}

\noindent \textbf{Acknowledgement}\\
It is a pleasure to thank Peter Zoller for numerous insightful suggestions that helped improve the manuscript. KRS further thanks Kazutaka Takahashi, Pablo Martínez-Azcona, and András Grabarits for useful discussions. This research was funded by the Luxembourg National Research Fund (FNR),  via the FNR-CORE Grant ``AQCQNET'' (C22/MS/17132054).

\bibliography{reference.bib}

% \newpage
\appendix
\newpage
\setcounter{secnumdepth}{2} % numbering in 

\subsection{Bound for $\beta(t)$}\label{apeendix:bound}
For a closed system under unitary dynamics, we have
\begin{equation}
\beta(t) = - A(t) = -\braket{| i[H_d, H_p] |}. 
\end{equation}
According to the Cauchy-Schwarz inequality, 
\begin{equation}
    \frac{1}{2} |\braket{| i[H_d, H_p] |} | \leq \Delta H_d \Delta H_p,\label{eq:cs-inequality}
\end{equation}
where $\Delta H= \sqrt{\braket{H^2}-\braket{H}^2}$ for an operator $H$. The variance is upper-bounded by the semi-norm $||H||$ of a Hermitian operator $H$ as $\Delta^2 H \leq ||H||/4$.
The semi-norm is defined as, $||H|| = M_H - m_H$, where $M_H (m_H)$ is the maximum (minimum) eigenvalue of $H$. Defining $n_p = \sqrt{||H_p||}$ and $n_d = \sqrt{||H_d||}$ and combining with Eq.~\eqref{eq:cs-inequality} we find
\begin{equation}
    \frac{n_p n_d}{4} \geq \Delta H_d \Delta H_p \geq \frac{1}{2} |\braket{| i[H_d, H_p] |} | = - \frac{1}{2} \beta(t),
\end{equation}
which introduces a lower bound for $\beta(t)$ as $\beta(t) \geq - \frac{n_p n_d}{2}$.

\subsection{Bound for $\Gamma_k(t)$}\label{apeendix:bound}
Consider the non-Markovian master equation, where we define
\begin{equation}
    \Gamma_k(t) = - \left(\braket{ P_k^\dagger H_p P_k } - \braket{H_p}\right)_t = \text{Tr}[\rho(t) ({ P_k^\dagger H_p P_k } - {H_p})]\label{appendix:na--qlc},
\end{equation}
and $ \Gamma_k(0) = \lambda_k$. The fidelity between the initial and the final reduced density matrices can be defined as 
\begin{equation}
    F(t) = {\text{Tr}[\rho(0) \rho(t)]},
\end{equation}
as the initial state is a known pure state, i.e., $\text{Tr}[\rho(0)^2] = 1$. Defining $ O_k =  { P_k^\dagger H_p P_k } - {H_p} $ leads to 
\begin{equation}
\begin{split}
    |\Gamma_k(t) - \Gamma_k(0)| \approx \text{Tr} [(\rho(t) - \rho(0))O_k] \leq ||O_k|| \cdot || \rho(t) - \rho(0) ||_1,
\end{split}
\end{equation}
where $||\cdot||$ is the operator norm (largest singular value), and $||\cdot||_1$ is the trace-norm. The trace distance is bounded by~
%\cite{Nielsen_Chuang_2010} 
\begin{equation}
    || \rho(t) - \rho(0) ||_1 \leq 2 \sqrt{1-F(t)}.
\end{equation}
Combining the two equations, we get a bound for the decay operators as 
\begin{equation}
    |\Gamma_k(t) - \Gamma_k(0) | = |\Gamma_k(t) - \lambda_k|  \leq 2 ~ ||O_k|| \cdot
 \sqrt{1-F(t)},\label{app:fidelity_bound}
\end{equation}
where the fidelity at short times can be calculated asymptotically (see Appendix.~\ref{apeendix:fid}).

However, a weaker but easily obtained bound can also be found using the Cauchy-Schwarz inequality for operators $ |\text{Tr} (A^{\dagger}B)|^2 \leq \text{Tr}(A^{\dagger}A) \text{Tr}(B^{\dagger}B)$. This leads to
\begin{equation}
    |\text{Tr}[\rho(t) ({ P_k^\dagger H_p P_k } )]| \leq \sqrt{\text{Tr}[\rho(t)^2]} ~ ||H_p||_2,
\end{equation}
which gives us the final inequality as
\begin{equation}
\begin{split}
    |\Gamma_k(t)| & \leq |\text{Tr}[\rho(t) ({ P_k^\dagger H_p P_k } )]| + || H_p || \\
    & \leq \sqrt{\mathcal{P}(t)} ~ ||H_p||_2 + || H_p || \\
    & \leq 2 ~ ||H_p||_2 ,
\end{split}
\end{equation}
where $\mathcal{P}(t) = {\text{Tr}[\rho(t)^2]} $ is the purity of the final state.

\begin{figure*}[t!]
    \centering
    \includegraphics[scale=0.32]{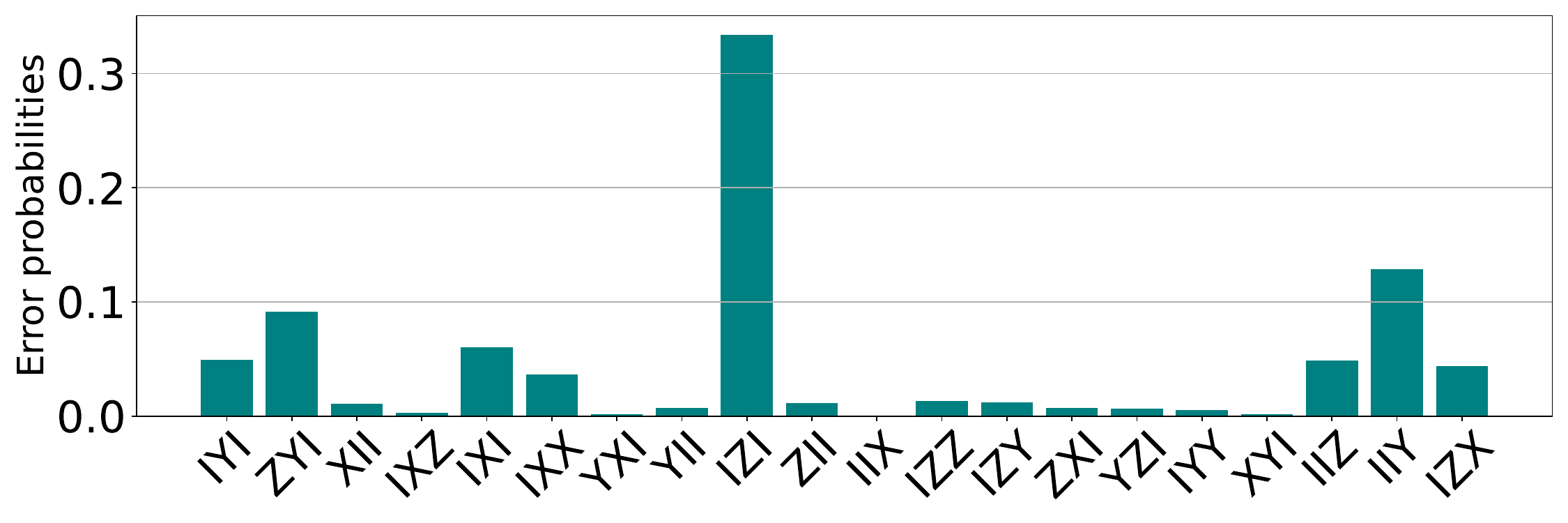} 
    \caption{Figure illustrating the randomly generated error probabilities used for the simulation of the spin-glass Hamiltonians (Fig.~\ref{TFIM_plot}) in the main text.}\label{second_err_probs}
\end{figure*}

\subsection{Fidelity for the short-time quantum decay}\label{apeendix:fid}
The bound defined in Eq.~\eqref{app:fidelity_bound} requires knowledge of the fidelity $F(t)$, and here we obtain an analytic expression for $F(t)$ in the short-time limit~\cite{Aurelia_adc_2017}. Its first derivative reads
\begin{equation}
\begin{split}
    \frac{d}{dt} F(t) & = \text{Tr} [\rho(0) \dot{\rho}(t) ] \\
    & = - i \text{Tr} \{ \rho(0) [H(t), \rho(t)] \} \\ & + \sum_{k\in \mathcal{K}}  \Gamma_k(t) \{ \text{Tr} [\rho(0) P_k \rho(t) P_k^{\dagger}] - \text{Tr} [\rho(0) \rho(t)] \}.
\end{split}
\end{equation}
At $t=0$, $F'(t)$ reduces to
\begin{equation}
\begin{split}
    F'(0) & =\lim_{t \rightarrow 0} \frac{d}{dt} F(t)\\ &  = - \sum_{k\in \mathcal{K}}  \Gamma_k(0) \{ \text{Tr}  [\rho(0) ^2 ] -  \text{Tr}  [\rho(0) P_k \rho(0) P_k^{\dagger} ] \}\\ & = - \sum_{k\in \mathcal{K}} \Gamma_k(0) \Delta P_k ^2,
\end{split}
\end{equation}
with the variance of the operator $\Delta P_k ^2 = \braket{P_k ^2} - \braket{P_k}^2$ and $ \Gamma_k(0) = \lambda_k$. Hence, in a short-time asymptotic expansion, the fidelity can be written as
\begin{equation}
\begin{split}
    F(t) & = 1 + F'(0) t + \mathcal{O} (t^2) \\
    & \approx 1 - t \sum_{k\in \mathcal{K}} \lambda_k \Delta P_k ^2.
\end{split}
\end{equation}

\subsection{Lyapunov controlled decoherence-free subspace}\label{lcdfs_app}
The decoherence-free subspace (DFS) method was introduced as an alternative method for error-free quantum computation~\cite{Lidar_Prl_dfs_1998}. This is an important concept, as the states in DFS evolve unitarily, even in the presence of decoherence. The Markovian dynamics representing the noise in the NISQ hardware is governed by Eq.~\eqref{rho_evolution} with the learned positive error rates, $\{\lambda\}$. Here, we add an extra Lyapunov control field $\gamma(t)$ and a control Hamiltonian $H_n$ to cancel out the dissipative part of the master equation. With the additional controls, the master equation becomes
\begin{equation}
    \frac{d \rho}{dt} = -i [H_p + \beta(t) H_d + \gamma(t) H_n, \rho] + \mathcal{D}_{\text{int}} [\rho],
\end{equation}
where $\mathcal{D}_{\text{int}} [\rho]$ is the stochastic Pauli noise channel of the NISQ device. For the new controlled Hamiltonian, Eq.~\eqref{open_system} reduces to
\begin{equation}
    \braket{\dot{H}_p} = \beta(t)  \braket{i[H_d, H_p]} + \gamma(t)  \braket{i[H_n, H_p]} + \text{Tr}(\mathcal{D}_{\text{int}} [\rho] H_p).
\end{equation}
Making a connection with~\cite{Wang2010}, we choose 
\begin{equation}
\begin{split}
    & \gamma(t) = - \frac{\text{Tr}(\mathcal{D}_{\text{int}} [\rho] H_p)}{\braket{i[H_n, H_p]}},\\
    & \beta(t) = - \braket{i[H_d, H_p]},
\end{split}
\end{equation}
such that $ \braket{\dot{H}_p} \leq 0$. There should always exist a $\gamma(t)$ unless $\braket{i[H_n, H_p]} = 0$.This approach can be called the Lyapunov-controlled decoherence-free subspace (LC-DFS).

Here we discuss the limitations of the LC-DFS compared to those of NAFQA. First, the choice of additional Lyapunov control Hamiltonians is restricted due to the condition $\braket{i[H_n, H_p]} \neq 0$. Second, the commutator $[H_n, H_p]$ may contain higher-order terms, which will be difficult to implement on quantum computers. For example, for the spin glass Hamiltonian considered in the main text, $H_p = \sum_{m<n}J_{mn}Z_m Z_n + \sum_l h_l Z_l$, and a two-body control Hamiltonian $H_n = \sum_{i<j} (Y_i Z_j + Z_i Y_j)$, the commutator $[H_n, H_p]$  contains one-body $X$, two-body $XZ, ZX$, and three-body $XZZ, ZZX$ operators. In this case, implementing the three-body operators on quantum computers is not straightforward, which further restricts the choice of control Hamiltonians and the application of the LC-DFS approach. 

\subsection{Amplitude and phase damping noise channels}
In the main text, we have considered the implementation of NAFQA with the stochastic Pauli noise (unital) channels. However, one can also study the behavior of the NAFQA algorithm beyond the stochastic Pauli noise channel, for instance, with the amplitude and phase damping channels, which are widely considered in OQS dynamics~\cite{Breuer_book}. The GKSL equation corresponding to the local amplitude and phase damping is given by
\begin{equation}
    \frac{d \rho}{dt} = -i [H, \rho] + \mathcal{D}_{\text{r}} [\rho]\label{ad_channel}. 
\end{equation}
The Lindblad dissipator takes the form
\begin{equation}
\begin{split}
    \mathcal{D}_{\text{r}} [\rho] & = \sum_{k = 1}^N \lambda_k \left(s_k^{(-)} \rho s_k^{(+)} - \frac{1}{2} \{s_k^{(+)} s_k^{(-)}, \rho\} \right) \\
    & + \sum_{k = 1}^N \frac{\lambda_k}{4} \left( Z_k \rho Z_k - \rho \right),
\end{split}
\end{equation}
which is a combination of the amplitude damping noise with decay rates $\lambda_k$ and the phase damping noise with rates $\lambda_k/4$. Here, $s_k^{(\pm)} = \frac{1}{2} \left( X_k \pm i Y_k \right)$, and $s_k^{(+)}  s_k^{(-)} = \ket{1}\bra{1}_k$ corresponds to the $k^{\text{th}}$ qubit. The condition for the single-qubit dephasing channel follows a procedure similar to that of the stochastic Pauli noise given in Eq.~\eqref{pauli_channel}. However, the Lyapunov condition for the amplitude-damping channel takes the form
\begin{equation}
\begin{split}
    & \frac{d}{dt} \braket{H_p} = \beta(t)  \braket{i[H_d, H_p]}\\
    & + \sum_{k\in \mathcal{K}} \Gamma_k(t) \left( \braket{ s_k^{(+)} H_p s_k^{(-)} } -\frac{1}{2} \braket{ \{H_p, s_k^{(+)} s_k^{(-)} \} } \right) \leq 0. \label{lyapunov_ad}
\end{split}
\end{equation}

The amplitude and phase damping channels can be implemented on quantum circuits with or without the help of ancilla qubits~\cite{Pleino_PRX_2023, Bassi_prr_2023}. Similarly to the feedback-based noise-assisted approach discussed in the Results, one can engineer the time-dependent decay rates based on the feedback law to use the amplitude and phase damping channels as a resource to assist quantum simulations.

\end{document}

%% file: shortcut.tex
  \newcommand{\rsym}[1]{\ensuremath{%
  \mathfrak{r}_{\mathrm{\ifx\@empty{}\else{#1}\fi}}}}

  \definecolor{cEdit}{rgb}{.17,.44,.76}